\documentstyle{article}

\newcommand{\noi}{\noindent}
\newcommand{\beq}[1]{\begin{equation}\label{#1}}
\newcommand{\eeq}{\end{equation}}
\newcommand{\bear}[1]{\begin{eqnarray}\label{#1}}
\newcommand{\ear}{\end{eqnarray}}
\newcommand{\nn}{\nonumber}

\catcode`\@=11 \@addtoreset{equation}{section}\catcode`\@=12
\newcommand{\be}{\begin{equation}}
\newcommand{\ee}{\end{equation}}
\newcommand{\ba}{\begin{eqnarray}}
\newcommand{\ea}{\end{eqnarray}}

\newcommand{\N}{ \mbox{\rm I$\!$N} }
\newcommand{\R}{ \mbox{\rm I$\!$R} }

\newcommand{\sign}{ \mbox{\rm sign} }
\newcommand{\e}{ \mbox{\rm e} }
\newcommand{\eps}{ \varepsilon }

\newcommand{\p}{\partial}
\newcommand{\btd}{\bigtriangledown}
\newcommand{\btu}{\bigtriangleup}
\newcommand{\tri}{\Delta}

\newcommand{\sums}{\sum\limits}
\newcommand{\const}{\mathop{\rm const}\nolimits}

\newcommand{\fnm}{\footnotemark}

\begin{document}

\noi
{\Large\bf
Multidimensional cosmological \\
and spherically symmetric solutions \\
with intersecting $p$-branes}

\bigskip

\noi
V.D. Ivashchuk\fnm[1] and V.N. Melnikov\fnm[1]

\bigskip

\noi
\fnm[1]{\small Center for Gravitation and Fundamental Metrology
VNIIMS, 3-1 M. Ulyanovoy Str. \\
Moscow, 117313, Russia e-mail: melnikov@rgs.phys.msu.su}

\bigskip

\noi
{\bf Abstract.}
Multidimensional model describing the "cosmological"
and/ or spherically symmetric configuration with  $(n+1)$
Einstein spaces in the theory with several scalar fields and forms is
considered.  When electro-magnetic composite
$p$-brane ansatz is adopted, $n$ "internal" spaces are Ricci-flat,
one space $M_0$ has a non-zero curvature, and all $p$-branes do not "live"
in $M_0$, a class of exact solutions is obtained if certain
block-orthogonality relations on p-brane vectors are imposed.
A subclass of spherically-symmetric solutions
containing non-extremal $p$-brane black holes is considered.
Post-Newtonian parameters are calculated and some examples are
considered.

\bigskip

\noi

\section{Introduction}
\setcounter{equation}{0}

The necessity of studying multidimensional models of gravitation
\cite{Mel2,Mel} is motivated by several reasons. First, the
main trend of modern physics is the unification of all known fundamental
physical interactions: electromagnetic, weak, strong and gravitational
ones. During last decades there was a significant progress in unifying
weak and electromagnetic interactions, some more modest achievements in
GUT, supersymmetric, string and superstring theories.

Now theories with membranes, p-branes and more vague M- and F-theories
\cite{HTW,Sc,Du,Vafa} are being
created and studied. Having no any definite successful theory of unification
now, it is desirable to study the common features of these theories and
their applications to solving basic problems of modern gravity.

Second, multidimensional gravitational models, as well as scalar-tensor
theories of gravity, are the theoretical framework for describing possible
temporal and range variations of fundamental physical constants \cite{3}.
These ideas originated from earlier papers of P.Dirac (1937) on relations
between fenomena of micro and macro worlds and up till now they are under
a thorough study both theoretically and experimentally.

Bearing in mind that multidimensional gravitational models are certain
generalizations of general relativity which is tested reliably for weak
fields up to 0,001 (they may be viewed as some effective scalar-tensor
theories in simple variants in four dimensions) it is quite natural to
inquire about their possible observational or experimental windows.
>From what we already know, among these windows are :

-- possible deviations from the Newton and Coulomb laws,

-- possible variations of the effective gravitational constant with
a time rate less than the Hubble one,

-- possible existence of monopole modes in gravitational waves,

-- different behaviour of strong field objects, such as multidimensional
black holes, wormholes and p-branes,

-- standard cosmological tests etc.

As no accepted unified model exists, in our approach we adopt simple, but
general from the point of view of number of dimensions, models based on
multidimensional Einstein equations with or without sources of different
nature:

-- cosmological constant,

-- perfect and viscous fluids,

-- scalar and electromagnetic fields,

-- plus their interactions,

-- fields of antisymmetric forms (related to p-branes) etc.

Our main objective was and is to obtain exact solutions (integrable models)
for these model self-consistent systems and then to analyze them in
cosmological, spherically and axially symmetric cases. In our view this
is a natural and most reliable way to study highly nonlinear systems.
It is done mainly within the Riemannian geometry. Some simple models in
integrable Weyle geometry and with torsion were studied also.

\subsection{Problem of Stability of G}

\subsubsection{Absolute G measurements}

The value of the Newton's gravitational constant $G$ as adopted by CODATA
in 1986 is based on Luther and Towler measurements of 1982.

Even at that time other existing on 100ppm level measurements deviated
>from this value more than their uncertainties \cite{6}. During last years
the situation, after very precise measurements of $G$ in Germany and New
Zealand, became much more vague. Their results deviate from the official
CODATA value from 600 ppm at minimal to 630 ppm at maximal values.

As it is seen from the most recent data announced in November 1998 at
the Cavendish conference in London the situation with terrestrail
absolute $G$ measurements is not improving.
The reported values for $G$ (in units of $10^{-11}$) and their estimated
error in ppm are as follows:
\begin{center}
\begin{tabular}{lll}
Fitzgerald and Armstrong & 6.6742 & 90 ppm \\
 & 6.6746 & 134 \\
Nolting et al. (Zurich) & 6.6749 & 210 \\
Meyer et al. (Wuppethal) & 6.6735 & 240 \\
Karagioz et al. (Moscow) & 6.6729 & 75 \\
Richman et al. & 6.683 & 1700 \\
Schwarz et al. & 6.6873 & 1400 \\
CODATA (1986, Luther) & 6.67259 & 128
\end{tabular}
\end{center}

This
means that either the limit of terrestrial accuracies is reached or we
have some new physics entering the measurement procedure \cite{7,8}.
First means that we should shift to space experiments to measure G \cite
{SEE} and second means that more thorough study of theories generalizing
Einstein's general relativity is necessary.

\subsubsection{Data on temporal variations of G}

Dirac's prediction based on his Large Numbers Hypothesis is $\dot{G}/G =
(-5)10^{-11}$ $year^{-1}$. Other hypotheses and theories, in particular
some scalar-tensor or multidimensional ones, predict these variations on
the level of $10^{-12}-10^{-13}$ per year. As to experimental or
observational
data, the results are rather nonconclusive. The most reliable ones are based
on Mars orbiters and landers (Hellings,1983) and on lunar laser ranging
(Muller et al., 1993; Williams et al., 1996). They are not better than
$10^{-12}$ per year \cite{9}. Here once more we see that there is a need for
corresponding theoretical and experimental studies. Probably, future space
missions to other planets will be a decisive step in solving the problem of
temporal variations of G and defining the fates of different theories which
predict them as the larger is the time interval between successive
measurements and, of course, the more precise they are, the more stringent
results will be obtained.

\subsubsection{Nonnewtonian interactions (EP and ISL tests)}

Nearly all modified theories of gravity and unified theories predict also
some deviations from the Newton law (ISL) or composite-dependant violation
of the Equivalence Principle (EP) due to an appearance of new possible
massive particles (partners) \cite{5}. Experimental data exclude the
existence
of these particles nearly at all ranges except less than millimeter and
also
at meters and hundreds of meters ranges. The most recent result in the range
of 20-500 m was obtained by Achilli et al \cite{10}. They found the positive
result for the deviation from the Newton law with the Yukawa potential
strength alpha between 0,13 and 0,25. Of course, these results need to be
verified in other independent experiments, probably in space ones.

\subsection{Multidimensional Models}

The history of multidimensional approach starts from the well-known papers
of T.K. Kaluza and O. Klein \cite{Kal,Kl} on 5-dimensional theories which
opened an interest (see \cite{DeS,Lee,Vl1,WeP1}) to investigations in
multidimensional gravity. These ideas were continued by P.Jordan \cite{J}
who suggested to consider the more general case $g_{55}\ne{\rm const}$
leading to the theory with an additional scalar field. The papers
\cite{Kal,Kl,J} were in some sense a source of inspiration for C. Brans
and R.H. Dicke in their well-known work on the scalar-tensor gravitational
theory \cite{BD}. After their work a lot of investigations were done using
material or fundamental scalar fields, both conformal and nonconformal
(see details in \cite{3}).

The revival of ideas of many dimensions started in 70th and continues now.
It is due completely to the development of unified theories. In the 70th an
interest to multidimensional gravitational models was stimulated mainly by:
i) the ideas of gauge theories leading to the non-Abelian generalization of
Kaluza-Klein approach  and by ii) supergravitational theories
\cite{CJS,SaSe}. In the 80th the supergravitational theories were "replaced"
by superstring models \cite{GrSW}. Now it is heated by expectations
connected
with overall M-theory or even some F-theory. In all these theories
4-dimensional gravitational models with extra fields were obtained from
some multidimensional model by a dimensional reduction based on the
decomposition of the manifold
$$
M=M^4\times M_{int},
$$
where $M^4$ is a 4-dimensional manifold and $M_{int}$ is some internal
manifold (mostly considered as a compact one).

The earlier papers on multidimensional cosmology dealt with multidimensional
Einstein equations and with a block-diagonal cosmological metric defined on
the manifold $M= \R \times M_0\times ...\times M_n$
of the form
$$
g=-dt\otimes dt+\sum_{r=0}^n a_r^2(t)  g^r
$$
where $(M_r,g^r)$ are Einstein spaces, $r=0,\dots,n$ \cite{FH}--\cite{Zh4}.
In some of them a cosmological constant and simple scalar fields were used
also \cite{BIMZ}.

In \cite{GRT,BL1,BL3,BIM1,IM2,BLP,IM3,GIM} the models with higher
dimensional "perfect-fluid" were considered. In these models pressures
(for any component) are proportional to the density
$$
p_r=\left(1-\frac{u_r}{d_r}\right)\rho,
$$
$r=0,\dots,n$, where $d_r$ is a dimension of $M_r$. Such models are
reduced to pseudo-Euclidean Toda-like systems with the Lagrangian
$$
L=\frac12G_{ij}\dot x^i\dot x^j-\sum_{k=1}^mA_k{\rm e}^{u_i^kx^i}
$$
and the zero-energy constraint $E=0$. In a classical case  exact solutions
with Ricci-flat $(M_r,g^r)$ for 1-component case were considered by many
authors (see, for example, \cite{Lor1,BO,BIM1,IM2,IM3,GIM,IM10} and
references therein). For the two component perfect-fluid there were
solutions with two curvatures, i.e.  $n=2$, when
$(d_1,d_2)=(2,8),(3,6),(5,5)$ \cite{GIM2} and corresponding non-singular
solutions from \cite{IM0}.  Among the solutions \cite{GIM2} there exists a
special class of Milne-type solutions.  Recently some interesting extensions
of 2-component solutions were obtained in \cite{GI}.

It should be noted that the pseudo-Euclidean Toda-like systems are not
well-studied yet. There exists a special class of equations of state that
gives rise to the Euclidean Toda models. First such solution was considered
in \cite{GIM} for the Lie algebra $a_2$. Recently the case of $a_n=sl(n+1)$
Lie algebras was considered and the solutions were expressed in terms of
a new elegant representation (obtained by Anderson) \cite{GKMR}.

The cosmological solutions may have regimes with: i) spontaneous and
dynamical compactifications; ii) Kasner-like and billiard behavior near the
singularity; iii) inflation and izotropization for large times (see, for
example, \cite{BIMZ,IM3}).

Near the singularity one can have an oscillating behavior like in
the well-known mixmaster  (Bianchi-IX) model. Multidimensional
generalizations of this model were considered by many authors (see, for
example, \cite{BelKh,BS,DHS,SzP}). In \cite{IvKM1,IvKM,IM6} the billiard
representation for multidimensional cosmological models near the
singularity was considered and the criterion for the volume of the
billiard to be finite was established in terms of illumination of the unit
sphere by point-like sources. For perfect-fluid this was considered in
detail in \cite{IM6}. Some interesting topics related to general
(non-homogeneous) situation were considered in \cite{KM}.

Multidimensional cosmological models have a generalization to the
case when the bulk and shear viscosity of the "fluid" is taken into account
\cite{GMT}. Some  classes of exact solutions were obtained, in particular
nonsingular cosmological solutions, generation of mass and entropy in the
Universe.

Multidimensional quantum cosmology based on the Wheeler-DeWitt (WDW)
equation
$$
\hat H\Psi=0,
$$
where $\Psi$ is the so-called "wave function of the universe", was
treated first in \cite{IMZ}, (see also \cite{Hal}). This equation was
considered for the vacuum case in \cite{IMZ} and integrated in a very
special situation of 2-spaces.
The WDW equation for the cosmological constant  and for the
"perfect-fluid" was investigated in  \cite{IMT,BIMZ} and \cite{IM3}
respectively.

Exact solutions in 1-component case were considered in detail
in \cite{IM10} (for perfect fluid). In \cite{Zh2} the multidimensional
quantum wormholes were suggested, i.e.  solutions with a special-type
behavior of the wave function (see \cite{HawP}).

These solutions were generalized to account for cosmological constant in
\cite{IMT,BIMZ} and to the perfect-fluid case in \cite{IM3,IM10}. In
\cite{IM6} the "quantum billiard" was obtained for
multidimensional WDW solutions near the
singularity.  It should be also noted that the "third-quantized"
multidimensional cosmological models were considered in several papers
\cite{Zh1,Hor,IM10}.  One may point out that in all cases when we had
classical cosmological solutions in many dimensions,
the  corresponding quantum cosmological solutions were found also.

Cosmological solutions are closely related to solutions with the spherical
symmetry. Moreover, the scheme of obtaining them is very similar to the
cosmological approach. The first multidimensional generalization of such
type was considered by D. Kramer \cite{Kr} and rediscovered by A.I. Legkii
\cite{Le}, D.J. Gross and M.J. Perry  \cite{GP} ( and also by Davidson and
Owen). In \cite{BrI} the Schwarzschild solution was generalized to
the case of $n$ internal Ricci-flat spaces and it was shown that black
hole configuration takes place when scale factors of internal spaces are
constants. In \cite{FIM2} an analogous generalization of the Tangherlini
solution \cite{Tan} was obtained. These solutions were also generalized to
the electrovacuum case \cite{FIM3,IM8,BM}. In \cite{BI,BM} multidimensional
dilatonic black holes were singled out. An interesting theorem was proved
in \cite{BM} that "cuts" all non-black-hole configurations as non-stable
under even monopole perturbations. In \cite{IM13} the extremely-charged
dilatonic black hole solution was generalized to
multicenter (Majumdar-Papapetrou)
case when the cosmological constant is non-zero.

We note that for $D =4$ the pioneering Majumdar-Papapetrou solutions
with conformal scalar field and electromagnetic field
were considered in \cite{Br}.

At present there exists a special interest to the so-called M-
and F-theories etc. \cite{HTW,Sc,Du,Vafa}.
These theories are "supermembrane" analogues of
superstring models \cite{GrSW} in $D=11,12$ etc. The low-energy limit of
these theories leads to models governed by the Lagrangian
$$
{\cal L}  = R[g]-
h_{\alpha\beta} g^{MN}\partial_{M}\varphi^\alpha\partial_{N}\varphi^\beta
-\sum_{a\in\Delta}\frac{\theta_a}{n_a!}\exp[2\lambda_{a}(\varphi)]
(F^a)^2,
$$
where $g$ is metric, $F^a=dA^a$ are forms of rank $F^a=n_a$, and
$\varphi^\alpha$ are scalar fields.

In \cite{IMC} it was shown that after dimensional reduction on the
manifold $M_0\times M_1\times\dots\times M_n$ and when the composite
$p$-brane ansatz is considered the problem is reduced to the gravitating
self-interacting $\sigma$-model with certain constraints imposed. For
electric $p$-branes see also \cite{IM0,IM,IMR}
(in  \cite{IMR} the composite electric case was considered). This
representation may be considered as a powerful tool for obtaining
different solutions with intersecting $p$-branes (analogs of membranes).
In \cite{IMC,IMBl}  the Majumdar-Papapetrou type solutions were obtained
(for non-composite electric case see \cite{IM0,IM} and
for composite electric case see \cite{IMR}).
These solutions correspond to Ricci-flat $(M_i,g^i)$,
$i=1,\dots,n$, and were generalized also to the case of Einstein internal
spaces \cite{IMC}.  Earlier some special classes of these solutions were
considered in \cite{PT,Ts1,GKT,AR,AEH,AIR}. The obtained solutions take
place, when certain orthogonality relations (on couplings parameters,
dimensions of "branes", total dimension) are imposed. In this situation a
class of cosmological and spherically-symmetric solutions was obtained
\cite{IMJ}. Special cases were also considered in \cite{LPX,BGIM,GrIM,BKR}.
The solutions with the horizon were considered in details in
\cite{CT,AIV,Oh,BIM,IMJ}. In \cite{BIM,Br1} some propositions related to i)
interconnection between the Hawking temperature and the singularity
behaviour, and ii) to multitemporal configurations were proved.

It should be noted that multidimensional and multitemporal generalizations
of the Schwarz\-schild and Tangherlini solutions were considered in
\cite{IM8,IM6I}, where the generalized  Newton's formulas in multitemporal
case were obtained.

We note also that  there exists a large variety of Toda solutions (open or
closed) when certain intersection rules are satisfied \cite{IMJ}.

In \cite{IMJ} (see also \cite{IMAJ})
the Wheeler-DeWitt equation was integrated for
intersecting $p$-branes in orthogonal case and corresponding classical
solutions were obtained also. A slightly different approach was suggested
in \cite{LMMP}. (For non-composite case see also \cite{GrIM}.)

In \cite{Cosm,I2} exact solutions for multidimensional
models with intersecting p-branes in case of static internal spaces were
obtained. They turned to be de Sitter or anti- de Sitter type. Generation
of the effective cosmological constant and inflation via p-branes was
demonstrated there. These solutions may be considered as an interesting
first step for a quantum description of low-energy
limits in different super-$p$-branes theories.

In this paper we continue our investigations of $p$-brane solutions
(see for example \cite{DKL,St,AIR} and references therein)
based on sigma-model approach   \cite{IMC,IM,IMR}
. (For pure gravitational sector see \cite{RZ,IM0}.)

Here we consider a cosmological and/or spherically symmetric case
, when all functions
depend upon one variable (time or radial variable). The model under
consideration contains several scalar fields and antisymmetric forms and is
governed by action (\ref{2.1i}).

The considered cosmological model contains some stringy cosmological
models (see for example \cite{LMPX}). It may be obtained (at classical
level) from multidimensional cosmological model with perfect fluid
\cite{IM3,GIM} as a special case.

Here we find a family of solutions depending on one variable
describing the (cosmological or spherically symmetric)
"evolution" of $(n+1)$
Einstein spaces in the theory with several scalar fields and forms.
 When an electro-magnetic composite
$p$-brane ansatz is adopted the field equations are
reduced to the equations for Toda-like system.

In the case when $n$ "internal" spaces are Ricci-flat,
one space $M_0$ has a non-zero curvature, and all $p$-branes do not "live"
in $M_0$, we find a family of solutions (Section 4) to the equations of
motion
(equivalent to equations for Toda-like Lagrangian with zero-energy
constraint \cite{IMJ}) if certain
block-orthogonality relations on $p$-brane
vectors $U^s$ are imposed. These solutions
generalize the solutions  from \cite{IMJ} with orthogonal
set of vectors $U^s$. A special class of "block-orthogonal" solutions
(with coinciding parameters $\nu_s$ inside blocks) was
considered earlier in \cite{Br1}.

Here we consider a subclass of spherically-symmetric solutions
(Sect. 5).
This subclass contains non-extremal $p$-brane black holes
for zero values of "Kas\-ner-like" parameters.
The relation for the Hawking temperature is presented (in
the black hole case).

We also calculate Post-Newtonian parameters $\beta$ and $\gamma$
(Eddington parameters) for the spherically-symmetric solutions
(Sect. 6). These parameters may
be useful for possible physical applications.

\section{\bf The model}
\setcounter{equation}{0}

Here like in \cite{IMC} we consider the model governed by the action
\bear{2.1i}
S =&& \frac{1}{2\kappa^{2}}
\int_{M} d^{D}z \sqrt{|g|} \{ {R}[g] - 2 \Lambda - h_{\alpha\beta}\;
g^{MN} \partial_{M} \varphi^\alpha \partial_{N} \varphi^\beta
\\ \nn
&& - \sum_{a \in \Delta}
\frac{\theta_a}{n_a!} \exp[ 2 \lambda_{a} (\varphi) ] (F^a)^2_g \}
+ S_{GH},
\ear
where $g = g_{MN} dz^{M} \otimes dz^{N}$ is the metric
($M,N =1, \ldots, D$), $\varphi=(\varphi^\alpha)\in \R^l$
is a vector from dilatonic scalar fields,
$(h_{\alpha\beta})$ is a non-degenerate symmetric
$l\times l$ matrix ($l\in \N$), $\theta_a = \pm 1$,
\beq{2.2i}
F^a =  dA^a =
\frac{1}{n_a!} F^a_{M_1 \ldots M_{n_a}}
dz^{M_1} \wedge \ldots \wedge dz^{M_{n_a}}
\eeq
is a $n_a$-form ($n_a \geq 1$) on a $D$-dimensional manifold $M$,
$\Lambda$ is cosmological constant and $\lambda_{a}$ is a $1$-form
on $\R^l$: $\lambda_{a} (\varphi) =\lambda_{a \alpha} \varphi^\alpha$,
$a \in \Delta$, $\alpha=1,\ldots,l$. In (\ref{2.1i}) we denote
$|g| = |\det (g_{MN})|$,
\beq{2.3i}
(F^a)^2_g =
F^a_{M_1 \ldots M_{n_a}} F^a_{N_1 \ldots N_{n_a}}
g^{M_1 N_1} \ldots g^{M_{n_a} N_{n_a}},
\eeq
$a \in \Delta$, where $\Delta$ is some finite set, and $S_{\rm GH}$ is the
standard Gibbons-Hawking boundary term \cite{GH}. In the models
with one time all $\theta_a =  1$  when the signature of the metric
is $(-1,+1, \ldots, +1)$.

The equations of motion corresponding to  (\ref{2.1i}) have the following
form
\bear{2.4i}
R_{MN} - \frac{1}{2} g_{MN} R  =   T_{MN} - \Lambda g_{MN},
\\
\label{2.5i}
{\btu}[g] \varphi^\alpha -
\sum_{a \in \Delta} \theta_a  \frac{\lambda^{\alpha}_a}{n_a!}
e^{2 \lambda_{a}(\varphi)} (F^a)^2_g = 0,
\\
\label{2.6i}
\nabla_{M_1}[g] (e^{2 \lambda_{a}(\varphi)}
F^{a, M_1 \ldots M_{n_a}})  =  0,
\ear
$a \in \Delta$; $\alpha=1,\ldots,l$.
In (\ref{2.5i}) $\lambda^{\alpha}_{a} = h^{\alpha \beta}
\lambda_{a \beta }$, where $(h^{\alpha \beta})$
is matrix inverse to $(h_{\alpha \beta})$.
In (\ref{2.4i})
\bear{2.7i}
T_{MN} =   T_{MN}[\varphi,g]
+ \sum_{a\in\Delta} \theta_a  e^{2 \lambda_{a}(\varphi)} T_{MN}[F^a,g],
\ear
where
\bear{2.8i}
T_{MN}[\varphi,g] =
h_{\alpha\beta}\left(\p_{M} \varphi^\alpha \p_{N} \varphi^\beta -
\frac{1}{2} g_{MN} \p_{P} \varphi^\alpha \p^{P} \varphi^\beta\right),
\\
T_{MN}[F^a,g] = \frac{1}{n_{a}!}\left[ - \frac{1}{2} g_{MN} (F^{a})^{2}_{g}
+ n_{a}  F^{a}_{M M_2 \ldots M_{n_a}} F_{N}^{a, M_2 \ldots M_{n_a}}\right].
\label{2.9i}
\ear
In (\ref{2.5i}), (\ref{2.6i}) ${\btu}[g]$ and ${\btd}[g]$
are Laplace-Beltrami and covariant derivative operators respectively
corresponding to  $g$.

Let us consider the manifold
\beq{2.10g}
M = \R  \times M_{0} \times \ldots \times M_{n}
\eeq
with the metric
\beq{2.11g}
g= w \e^{2{\gamma}(u)} du \otimes du +
\sum_{i=0}^{n} \e^{2\phi^i(u)} g^i ,
\eeq
where $w=\pm 1$, $u$ is a distinguished coordinate which, by
convention, will be called ``time";
$g^i  = g^i_{m_{i} n_{i}}(y_i) dy_i^{m_{i}} \otimes dy_i^{n_{i}}$
is a metric on $M_{i}$  satisfying the equation
\beq{2.12g}
R_{m_{i}n_{i}}[g^i ] = \xi_{i} g^i_{m_{i}n_{i}},
\eeq
$m_{i},n_{i}=1,\ldots,d_{i}$; $d_{i} = \dim M_i$, $\xi_i= \const$,
$i=0,\dots,n$; $n \in \N$. Thus, $(M_i,g^i)$ are Einstein spaces.
The functions $\gamma,\phi^i$: $(u_-,u_+)\to\R$ are smooth.

Each manifold $M_i$ is assumed to be oriented and connected,
$i = 0,\ldots,n$. Then the volume $d_i$-form
\beq{2.13g}
\tau_i  = \sqrt{|g^i(y_i)|}
\ dy_i^{1} \wedge \ldots \wedge dy_i^{d_i},
\eeq
and the signature parameter
\beq{2.14g}
\eps(i)  = \sign \det (g^i_{m_{i}n_{i}}) = \pm 1
\eeq
are correctly defined for all $i=0,\ldots,n$.

Let
\beq{2.15g}
\Omega_0 = \{ \emptyset, \{ 0 \}, \{ 1 \}, \ldots, \{ n \},
\{ 0, 1 \}, \ldots, \{ 0, 1,  \ldots, n \} \}
\eeq
be a set of all subsets of
\beq{2.25n}
I_0\equiv\{ 0, \ldots, n \}.
\eeq
Let $I = \{ i_1, \ldots, i_k \} \in \Omega_0$, $i_1 < \ldots < i_k$.
We define a form
\beq{2.17i}
\tau(I) \equiv \tau_{i_1}  \wedge \ldots \wedge \tau_{i_k},
\eeq
of rank
\beq{2.19i}
d(I) \equiv  \sum_{i \in I} d_i,
\eeq
and a corresponding $p$-brane submanifold
\beq{2.18i}
M_{I} \equiv M_{i_1}  \times  \ldots \times M_{i_k},
\eeq
where $p=d(I)-1$ (${\rm dim M_{I}} = d(I)$).
We also define $\eps$-symbol
\beq{2.19e}
\eps(I) \equiv  \eps(i_1) \ldots \eps(i_k).
\eeq
For $I = \emptyset$ we put  $\tau(\emptyset) = \eps(\emptyset) = 1$,
 $d(\emptyset) = 0$.

For fields of forms we adopt the following "composite electro-magnetic"
ansatz
\beq{2.27n}
F^a=
\sum_{I\in\Omega_{a,e}}{\cal F}^{(a,e,I)}+\sum_{J\in\Omega_{a,m}}{\cal
F}^{(a,m,J)},
\eeq
where
\bear{2.28n}
{\cal F}^{(a,e,I)}=d\Phi^{(a,e,I)}\wedge\tau(I), \\ \label{2.29n}
{\cal F}^{(a,m,J)}=\e^{-2\lambda_a(\varphi)}
*\left(d\Phi^{(a,m,J)}\wedge\tau(J)\right),
\ear
$a\in\tri$, $I\in\Omega_{a,e}$, $J\in\Omega_{a,m}$ and
\beq{2.29nn}
\Omega_{a,e},\Omega_{a,m}\subset \Omega_0.
\eeq
(For empty $\Omega_{a,v}=\emptyset$, $v=e,m$, we put $\sums_\emptyset=0$ in
(\ref{2.27n})). In (\ref{2.29n}) $*=*[g]$ is the Hodge operator on $(M,g)$.

For the potentials in (\ref{2.28n}), (\ref{2.29n}) we put
\beq{2.28nn}
\Phi^s=\Phi^s(u),
\eeq
$s\in S$, where
\beq{6.39i}
S=S_e \sqcup S_m,  \qquad
S_v \equiv \sqcup_{a\in\tri}\{a\}\times\{v\}\times\Omega_{a,v},
\eeq
$v=e,m$. Here $\sqcup$ means the union of non-intersecting sets.
The set  $S$ consists of elements
$s=(a_s,v_s,I_s)$, where $a_s \in \tri$, $v_s = e,m$ and
$I_s \in \Omega_{a,v_s}$ are "color", "electro-magnetic" and
"brane" indices, respectively.

For dilatonic scalar fields we put
\beq{2.30n}
\varphi^\alpha=\varphi^\alpha(u),
\eeq
$\alpha=1,\dots,l$.

>From  (\ref{2.28n})  and (\ref{2.29n}) we obtain
the relations between dimensions of $p$-brane
worldsheets and ranks of forms
\bear{2.d1}
d(I) = n_a - 1,  \quad I \in \Omega_{a,e},
\\ \label{2.d2}
d(J) = D - n_a - 1,  \quad J \in \Omega_{a,m},
\ear
in electric and magnetic cases respectively.

\section{\bf $\sigma$-model representation}
\setcounter{equation}{0}

Here, like in \cite{IMJ}, we impose a restriction on
$p$-brane configurations, or, equivalently, on $\Omega_{a,v}$.
We assume that the energy momentum tensor $(T_{MN})$ has a block-diagonal
structure (as it takes place for $(g_{MN})$).
Sufficient  restrictions on $\Omega_{a,v}$ that guarantee
a block-diagonality of $(T_{MN})$ are presented in Appendix 1.

It follows from \cite{IMC} (see Proposition 2 in \cite{IMC}) that the
equations of motion (\ref{2.4i})--(\ref{2.6i}) and the Bianchi
identities
\beq{2.b}
d{\cal F}^s=0, \quad s\in S
\eeq
for the field configuration (\ref{2.11g}), (\ref{2.27n})--(\ref{2.29n}),
\ref{2.28nn}, (\ref{2.30n}) with the restrictions (\ref{B.31in}),
(\ref{B.17in}) (from Appendix 1)
imposed are equivalent to equations of motion for $\sigma$-model
with the action
\bear{2.25gn}
S_{\sigma} = \frac{\mu_{*}}2
\int du {\cal N} \biggl\{G_{ij}\dot\phi^i\dot\phi^j
+h_{\alpha\beta}\dot\varphi^{\alpha}\dot\varphi^{\beta} \\ \nonumber
+\sum_{s\in S}\eps_s\exp[-2U^s(\phi,\varphi)](\dot\Phi^s)^2
-2{\cal N}^{-2}V(\phi)\biggr\},
\ear
where $\dot x\equiv dx/du$,
\beq{2.27gn}
V = {V}(\phi) = -w\Lambda\e^{2\gamma_0(\phi)}+
\frac w2\sum_{i =0}^{n} \xi_i d_i \e^{-2 \phi^i + 2 {\gamma_0}(\phi)}
\eeq
is the potential with
\beq{2.24gn}
\gamma_0(\phi)
\equiv\sum_{i=0}^nd_i\phi^i,  \label{2.32g}
\eeq
and
\beq{2.24gn1}
{\cal N}=\exp(\gamma_0-\gamma)>0
\eeq
is the lapse function,
\bear{2.u}
 U^s = U^s(\phi,\varphi)= -\chi_s\lambda_{a_s}(\varphi) +
\sum_{i\in I_s}d_i\phi^i, \\ \label{2.e}
\eps_s=(-\eps[g])^{(1-\chi_s)/2}\eps(I_s)\theta_{a_s}
\ear
for $s=(a_s,v_s,I_s)\in S$, $\eps[g]= \sign \det (g_{MN})$,
(more explicitly (\ref{2.e}) reads $\eps_s=\eps(I_s) \theta_{a_s}$ for
$v_s = e$ and $\eps_s=-\eps[g] \eps(I_s) \theta_{a_s}$, for
$v_s = m$)
\bear{2.x1}
\chi_s=+1, \quad v_s=e; \\ \label{2.x2}
\chi_s=-1, \quad v_s=m,
\ear
and
\beq{2.c}
G_{ij}=d_i\delta_{ij}-d_id_j
\eeq
are components of the "pure cosmological" minisupermetric; $i,j=0,\dots,n$
\cite{IMZ}.

In the electric case $({\cal F}^{(a,m,I)}=0)$ for finite internal space
volumes $V_i$ the action (\ref{2.25gn}) coincides with the
action (\ref{2.1i}) if
$\mu_{*} =-w/\kappa_0^2$, $\kappa^{2} = \kappa^{2}_0 V_0 \ldots V_n$.

Action (\ref{2.25gn}) may be also written in the form
\beq{2.31n}
S_\sigma=\frac{\mu_*}{2}\int du{\cal N}\left\{
{\cal G}_{\hat A\hat B}(X)\dot X^{\hat A}\dot X^{\hat B}-
2{\cal N}^{-2}V(X)\right\},
\eeq
where $X = (X^{\hat A})=(\phi^i,\varphi^\alpha,\Phi^s)\in{\bf
R}^{N}$, and minisupermetric \beq{2.31nn} {\cal G}={\cal G}_{\hat
A\hat B}(X)dX^{\hat A}\otimes dX^{\hat B} \eeq on minisuperspace
\beq{2.31m}
{\cal M}=\R^{N}, \quad   N = n+1+l+|S|
\eeq
($|S|$ is the number of elements in $S$) is defined by the relation
\beq{2.33n}
({\cal G}_{\hat A\hat B}(X))=\left(\begin{array}{ccc}
G_{ij}&0&0\\[5pt]
0&h_{\alpha\beta}&0\\[5pt]
0&0&\eps_s\e^{-2U^s(X)}\delta_{ss'}
\end{array}\right).
\eeq

The minisuperspace metric (\ref{2.31nn}) may be also written in the form
\beq{2.34n}
{\cal G}=\bar G+\sum_{s\in S}\eps_s\e^{-2U^s(x)}d\Phi^s\otimes d\Phi^s,
\eeq
where $x=(x^A)=(\phi^i,\varphi^\alpha)$,
\bear{2.35n}
\bar G=\bar G_{AB}dx^A\otimes dx^B=G_{ij}d\phi^i\otimes d\phi^j+
h_{\alpha\beta}d\varphi^\alpha\otimes d\varphi^\beta, \\ \label{2.36n}
(\bar G_{AB})=\left(\begin{array}{cc}
G_{ij}&0\\
0&h_{\alpha\beta}
\end{array}\right),
\ear
$U^s(x)=U_A^sx^A$ is defined in  (\ref{2.u}) and
\beq{2.38n}
(U_A^s)=(d_i\delta_{iI_s},-\chi_s\lambda_{a_s\alpha}).
\eeq
Here
\beq{2.39n}
\delta_{iI}\equiv\sum_{j\in I}\delta_{ij}=\begin{array}{ll}
1,&i\in I\\
0,&i\notin I
\end{array}
\eeq
is an indicator of $i$ belonging to $I$. The potential (\ref{2.27gn})
reads
\beq{2.40n}
V=(-w\Lambda)\e^{2U^\Lambda(x)}+\sum_{j=0}^n\frac w2\xi_jd_j
\e^{2U^j(x)},
\eeq
where
\bear{2.41n}
U^j(x)=U_A^jx^A=-\phi^j+\gamma_0(\phi), \\ \label{2.42n}
U^\Lambda(x)=U_A^\Lambda x^A=\gamma_0(\phi), \\ \label{2.43n}
(U_A^j)=(-\delta_i^j+d_i,0), \\ \label{2.44n}
(U_A^\Lambda)=(d_i,0).
\ear

The integrability of the Lagrange system (\ref{2.31n}) depends
upon the scalar products of co-vectors $U^\Lambda$, $U^j$, $U^s$
corresponding to $\bar G$:
\beq{2.45n}
(U,U')=\bar G^{AB}U_AU'_B,
\eeq
where
\beq{2.46n}
(\bar G^{AB})=\left(\begin{array}{cc}
G^{ij}&0\\
0&h^{\alpha\beta}
\end{array}\right)
\eeq
is matrix inverse to (\ref{2.36n}). Here (as in \cite{IMZ})
\beq{2.47n}
G^{ij}=\frac{\delta^{ij}}{d_i}+\frac1{2-D},
\eeq
$i,j=0,\dots,n$. These
products have the following form
\bear{2.48n}
(U^i,U^j)=\frac{\delta_{ij}}{d_j}-1,
\\ \label{2.50n}
(U^\Lambda,U^\Lambda)=-\frac{D-1}{D-2}, \\ \label{2.51n}
(U^s,U^{s'})=q(I_s,I_{s'})+ \chi_s \chi_{s'}
\lambda_{a_s}\cdot\lambda_{a_{s'}}, \\ \label{2.52n}
(U^s,U^i)=-\delta_{iI_s},
\ear
where $s=(a_s,v_s,I_s)$, $s'=(a_{s'},v_{s'},I_{s'})\in S$,
\bear{2.54n}
q(I,J)\equiv d(I\cap J)+\frac{d(I)d(J)}{2-D}, \\ \label{2.55n}
\lambda_a\cdot\lambda_b\equiv\lambda_{a\alpha}\lambda_{b\beta}
h^{\alpha\beta}.
\ear
Relations (\ref{2.48n})-(\ref{2.50n})  were found in
\cite{GIM}  and (\ref{2.51n})   in \cite{IMC}.

\section{\bf Cosmological and spherically \protect\\ symmetric solutions}
\setcounter{equation}{0}

Here we put the following restrictions on the parameters of the model
\beq{5.1n}
{\bf (i)}\qquad \Lambda=0,
\eeq
i.e. the cosmological constant is zero,
\beq{5.2n}
{\bf (ii)}\qquad \xi_0\ne0, \quad \xi_1=\dots=\xi_n=0,
\eeq
i.e. one space is curved and others are Ricci-flat,
\beq{5.3n}
{\bf (iii)}\qquad 0\notin I_s, \quad \forall s=(a_s,v_s,I_s)\in S,
\eeq
i.e. all "brane" manifolds $M_{I_s}$ (see (\ref{2.18i})) do not
contain $M_0$.

We also impose a block-orthogonality restriction on the
set of vectors $(U^s, s \in S)$.
Let
\beq{5.3an}
S=S_1\sqcup\dots\sqcup S_k,
\eeq
$S_i\ne\emptyset$, $i=1,\dots,k$, and
\beq{5.4n}
{\bf (iv)} \
(U^s,U^{s'})=d(I_s\cap I_{s'})+\frac{d(I_s)d(I_{s'})}{2-D}+
\chi_s\chi_{s'}\lambda_{a_s\alpha}\lambda_{a_{s'}\beta}h^{\alpha\beta}=0,
\eeq
for all $s =(a_s,v_s,I_s) \in S_i$,
$s'=(a_{s'},v_{s'},I_{s'}) \in S_j$,
$i\ne j$; $i,j=1,\dots,k$. Relation
(\ref{5.3an}) means that the set $S$ is a union of $k$ non-intersecting
(non-empty) subsets $S_1,\dots,S_k$. According to (\ref{5.4n}) the set of
vectors $(U^s,s\in S)$ has a block-orthogonal structure with respect to
the scalar product (\ref{2.45n}), i.e. it splits into $k$ mutually
orthogonal blocks $(U^s,s\in S_i)$, $i=1,\dots,k$.

>From {\bf (i)}, {\bf (ii)}  we get for the potential (\ref{2.40n})
\beq{5.6n}
V=\frac12w\xi_0d_0\e^{2U^0(x)},
\eeq
where
\beq{5.7n}
(U^0,U^0)=\frac1{d_0}-1<0
\eeq
(see (\ref{2.48n})).

>From {\bf (iii)} and (\ref{2.52n}) we get
\beq{5.8n}
(U^0,U^{s})=0
\eeq
for all $s\in S$.
Thus, the set of co-vectors $U^0$, $U^s, s \in S$ (belonging
to dual space $(\R^{n+1+l})^*\simeq\R^{n+1+l}$)  has also a
block-orthogonal structure with respect
to the scalar product (\ref{2.45n}).

Here we fix the time gauge as follows
\beq{4.1n}
\gamma  = \gamma_0,  \quad  {\cal N} = 1,
\eeq
i.e the harmonic time gauge is used.
Then we obtain the
Lagrange system with the Lagrangian
\beq{3.14r}
L=\frac{\mu_{*}}{2} {\cal G}_{\hat A\hat B}(X)
\dot X^{\hat A}\dot X^{\hat B}- \mu_{*} V
\eeq
and the energy constraint
\beq{3.15r}
E=\frac{\mu_{*}}{2} {\cal G}_{\hat A\hat B}(X)
\dot X^{\hat A}\dot X^{\hat B}+\mu_{*}V= 0.
\eeq

Here we will integrate the Lagrange equations corresponding to the
Lagrangian (\ref{3.14r}) with the energy-constraint (\ref{3.15r}) and
hence we will find classical exact solutions when the restrictions
(\ref{B.31in}), (\ref{B.17in}) from Appendix 1
are imposed.

The problem of integrability may be simplified if we integrate the Maxwell
equations (for $s\in S_e$) and Bianchi identities (for $s\in S_m$):
\bear{5.29n}
\frac d{du}\left(\exp(-2U^s)\dot\Phi^s\right)=0
\Longleftrightarrow
\dot\Phi^s=Q_s \exp(2U^s),
\ear
where $Q_s$ are constants, $s \in S$.

Let
\bear{5.30n}
Q_s \ne 0,
\ear
for all $s \in S$.

For fixed $Q=(Q_s, s \in S)$ the Lagrange equations for the Lagrangian
(\ref{3.14r}) corresponding to $(x^A)=(\phi^i,\varphi^\alpha)$,
when equations (\ref{5.29n}) are substituted are equivalent to the Lagrange
equations for the Lagrangian
\beq{5.31n}
L_Q=\frac12\bar G_{AB}\dot x^A\dot x^B-V_Q,
\eeq
where
\beq{5.32n}
V_Q=V+\frac12\sum_{s\in S}\eps_sQ_s^2\exp[2U^s(x)],
\eeq
$(\bar G_{AB})$ and $V$ are defined in (\ref{2.36n}) and (\ref{5.6n})
respectively. The zero-energy constraint (\ref{3.15r}) reads
\beq{5.33n}
E_Q=\frac12\bar G_{AB}\dot x^A\dot x^B+V_Q=0.
\eeq

When the conditions {\bf (i)}--{\bf (iv)} are satisfied  exact solutions
to Lagrange equations corresponding to
(\ref{5.31n}) with the potential (\ref{5.32n}) and $V$ from
(\ref{5.6n}) could be readily obtained using the relations from Appendix 2.

The solutions read:
\beq{5.34n}
x^A(u)= - \frac{U^{0A}}{(U^0,U^0)}\ln |f_0(u)| -
\sum_{s\in S} \eta_s \nu_s^2 U^{sA} \ln |f_s(u)| + c^A u +
\bar{c}^A.
\eeq

Functions $f_0$ and $f_s$
in (\ref{5.34n}) are the following:
\bear{5.35no}
f_0(u)=
\left|\xi_0(d_0-1) \right|^{1/2}
s(u-u_0, w \xi_0,C_0) =
\\ \label{5.36o}
\left|\frac{\xi_0(d_0-1)}{C_0}\right|^{1/2}
\sinh(\sqrt{C_0}(u-u_0)), \  C_0>0,   \  \xi_0w>0; \\ \label{5.36n}
\left|\frac{\xi_0(d_0-1)}{C_0}\right|^{1/2}
\sin(\sqrt{|C_0|}(u-u_0)), \  C_0<0, \  \xi_0w>0; \\ \label{5.37n}
\left|\frac{\xi_0(d_0-1)}{C_0}\right|^{1/2}
\cosh(\sqrt{C_0}(u-u_0)), \  C_0>0, \  \xi_0w<0; \\ \label{5.38n}
\left|\xi_0(d_0-1)\right|^{1/2}
(u-u_0), \  C_0=0, \  \xi_0w>0,
\ear
and
\bear{5.39no}
f_s(u)=
\frac{|Q_s|}{|\nu_s|}
s(u-u_s, -\eta_s\eps_s,C_s) = \\ \label{5.39n}
\frac{|Q_s|}{|\nu_s||C_s|^{1/2}}\sinh(\sqrt{C_s}(u-u_s)), \;
C_s>0, \; \eta_s\eps_s<0; \\ \label{5.40n}
\frac{|Q_s|}{|\nu_s||C_s|^{1/2}}\sin(\sqrt{|C_s|}(u-u_s)), \;
C_s<0, \; \eta_s\eps_s<0; \\ \label{5.41n}
\frac{|Q_s|}{|\nu_s||C_s|^{1/2}}\cosh(\sqrt{C_s}(u-u_s)), \;
C_s>0, \; \eta_s\eps_s>0; \\ \label{5.42n}
\frac{|Q^s|}{|\nu_s|}(u-u_s), \; C_s=0, \; \eta_s\eps_s<0,
\ear
where $C_0$, $C_s$, $u_0$, $u_s$ are constants, $s \in S$.
The function $s(u, \xi,C)$ is defined in Appendix 2.

The parameters $\eta_s = \pm 1$, $\nu_s \neq 0$, $s\in S$, satisfy
the relations
\bear{5.42p}
\sum_{s' \in S} (U^s, U^{s'}) \eta_{s'} \nu_{s'}^2 = 1,
\ear
for all $s \in S$, with scalar products $(U^s, U^{s'})$ defined in
(\ref{2.51n}).

The constants $C_s$, $u_s$
are coinciding inside blocks:
\bear{5.42o}
u_s = u_{s'}, \qquad C_s = C_{s'},
\ear
$s,s' \in S_i$, $i = 1, \ldots, k$
(see relation (\ref{A.11}) from Appendix 2).
The ratios $\eps_s Q_s^2/(\eta_{s} \nu_{s}^2)$ are
also coinsiding inside blocks, or, equivalently,
\bear{5.42r}
\eps_s \eta_{s} = \eps_{s'} \eta_{s'}
\\ \label{5.42s}
\frac{Q_s^2}{\nu_s^2} = \frac{Q_{s'}^2}{\nu_{s'}^2},
\ear
$s,s' \in S_i$, $i = 1, \ldots, k$.
Here we used the relations (\ref{5.7n}), (\ref{5.8n}).

The contravariant components $U^{rA}= \bar
G^{AB} U^r_B$ are \cite{IMJ}
\bear{5.43n}
U^{0i}=-\frac{\delta_0^i}{d_0}, \quad U^{0\alpha}=0,
\\ \label{4.8n}
U^{si}= G^{ij}U_j^s= \delta_{iI_s}-\frac{d(I_s)}{D-2}, \quad
U^{s\alpha}= - \chi_s \lambda_{a_s}^\alpha.
\ear

Using (\ref{5.34n}), (\ref{5.7n}), (\ref{4.8n}) and
(\ref{5.43n}) we obtain
\beq{5.44n}
\phi^i=\frac{\delta_0^i}{1-d_0}\ln |f_0|
- \sum_{s\in S} \eta_s\nu_s^2\left(\delta_{iI_s}-\frac{d(I_s)}{D-2}\right)
\ln |f_s| + c^iu+\bar c^i,
\eeq
and
\beq{5.46n}
\varphi^\alpha=\sum_{s\in S}\eta_s\nu_s^2\chi_s\lambda_{a_s}^\alpha
\ln |f_s|+c^\alpha u+\bar c^\alpha,
\eeq
$\alpha=1,\dots,l$.

Vectors $c=(c^A)$ and $\bar c=(\bar c^A)$ satisfy the linear constraint
relations (see (\ref{A.18}) in  Appendix 2)
\bear{5.47n}
U^0(c)= U^0_A c^A = -c^0+\sum_{j=0}^nd_jc^j=0, \\ \label{5.48n}
U^0(\bar c)= U^0_A \bar c^A =
-\bar c^0+\sum_{j=0}^nd_j\bar c^j=0, \\ \label{5.49n}
U^s(c)= U^s_A c^A=
\sum_{i\in I_s}d_ic^i-\chi_s\lambda_{a_s\alpha}c^\alpha=0,
\\ \label{5.50n}
U^s(\bar c)=  U^s_A \bar c^A=
\sum_{i\in I_s}d_i\bar c^i-
\chi_s\lambda_{a_s\alpha}\bar c^\alpha=0,
\ear
$s\in S$. The (\ref{2.24gn}) reads
\beq{5.51n}
\gamma_0(\phi) = \frac{d_0}{1-d_0}\ln |f_0|+
\sum_{s\in S}\frac{d(I_s)}{D-2}\eta_s\nu_s^2\ln |f_s| +
c^0u+\bar c^0.
\eeq

The zero-energy constraint reads (see Appendix 2)
\beq{5.53n}
E=E_0+\sum_{s\in S}E_s+ \frac12 \bar G_{AB}c^Ac^B=0,
\eeq
where $E_0 = C_0(U^0,U^0)^{-1}/2$, $E_s = C_s (\eta_s \nu_s^2)/2$.
Using relations
(\ref{2.c}), (\ref{2.36n}),
(\ref{5.7n}) and (\ref{5.47n})
we rewrite (\ref{5.53n}) as
\beq{5.55n}
C_0\frac{d_0}{d_0-1}=\sum_{s\in S} C_s\nu_s^2\eta_s+
h_{\alpha\beta}c^\alpha c^\beta+\sum_{i=1}^n d_i(c^i)^2+
\frac1{d_0-1}\left(\sum_{i=1}^nd_ic^i\right)^2.
\eeq

>From relation
\beq{5.56n}
\exp(2 U^s) =f_s^{-2},
\eeq
following from  (\ref{5.4n}), (\ref{5.8n}), (\ref{5.34n}),
(\ref{5.49n}) and (\ref{5.50n}) we get for
electric-type forms (\ref{2.28n})
\beq{5.57n}
{\cal F}^s=Q_s f_s^{-2}du\wedge\tau(I_s),
\eeq
$s\in S_e$, and for magnetic-type forms (\ref{2.29n})
\beq{5.58n}
{\cal F}^s=\e^{-2\lambda_a(\varphi)}
*\left[Q_s f_s^{-2} du \wedge\tau(I_s)\right] =
\bar Q_s \tau(\bar I_s),
\eeq
$s\in S_m$, where  $\bar Q_s=Q_s\eps(I_s)\mu(I_s)w$
and $\mu(I) =\pm1$ is defined by the relation
$\mu(I) du \wedge \tau(I_0)=\tau(\bar I)\wedge du\wedge\tau(I)$.
The relation (\ref{5.58n}) follows from the formula
(5.26) from \cite{IMC} (for $\gamma=\gamma_0$).

Relations for the metric follows from (\ref{5.44n})
and (\ref{5.51n})
\bear{5.63n}
g= \biggl(\prod_{s\in S}
[f_s^2(u)]^{\eta_s d(I_s)\nu_s^2/(D-2)}\biggr)
\biggl\{[f_0^2(u)]^{d_0/(1-d_0)}\e^{2c^0u+2\bar c^0}\\ \nn
\times[wdu\otimes du+f_0^2(u)g^0]+
\sum_{i =1}^{n} \Bigl(\prod_{s\in S }
[f_s^2(u)]^{-\eta_s\nu_s^2 \delta_{i I_s} }\Bigr)\e^{2c^iu+2\bar
c^i}g^i\biggr\}.
\ear

Thus, here we obtained the "block-orthogonal" generalization of the
solution from \cite{IMJ}.
This  solution describes the evolution of $n+1$ spaces
$(M_0,g_0),\dots,  \\
(M_n,g_n)$, where $(M_0,g_0)$ is an Einstein
space of non-zero curvature, and
$(M_i,g^i)$ are "internal" Ricci-flat spaces, $i=1,\dots,n$; in the
presence of several scalar fields and forms.
The solution is presented
by relations  (\ref{5.46n}), (\ref{5.57n})-(\ref{5.63n})
with the functions $f_0$, $f_s$  defined in
(\ref{5.35no})--(\ref{5.42n}) and the relations on the parameters of
solutions $c^A$, $\bar c^A$ $(A=i,\alpha)$, $C_0$, $C_s$, $u_s$, $Q_s$ ,
$\eta_s$, $\nu_s$ ($s\in S$) imposed in
(\ref{5.42p})--(\ref{5.42s}), (\ref{5.47n})--(\ref{5.50n}),
(\ref{5.55n}),  respectively.

This solution describes a set of charged (by forms) overlapping
$p$-branes ($p_s=d(I_s)-1$, $s\in S$) "living" on submanifolds
(isomorphic to)
$M_{I_s}$ (\ref{2.18i}), where the sets $I_s$ do not contain $0$,
i.e. all $p$-branes live in "internal" Ricci-flat spaces.

The solution is valid if the dimensions of $p$-branes and dilatonic
coupling vector satisfy the relations (\ref{5.4n}).
In "orthogonal" non-composite case these solutions were considered
in \cite{GrIM,BGIM} (electric case) and \cite{BIM}
(electro-magnetic case). For $n = 1$ see also \cite{LMPX,LPX}.
In block-orthogonal (non-composite) case
a special class of solutions with  $\nu_s^2$ coinciding inside blocks
was considered earlier in \cite{Br1}.

\section{Spherically symmetric and black hole \protect\\ solutions}

Here we consider the spherically symmetric case
\beq{5.64}
w = 1, \quad M_0 = S^{d_0}, \quad g^0 = d \Omega^2_{d_0},
\eeq
where $d \Omega^2_{d_0}$ is the canonical metric on a unit sphere
$S^{d_0}$, $d \geq 2$. We also assume that
\beq{5.64a}
M_1 = \R, \quad
g^1 = - dt \otimes dt
\eeq
(here $M_1$ is a time manifold) and
\beq{5.65}
1 \in I_s, \quad \forall s \in S,
\eeq
i. e. all p-branes have a common time direction $t$.

For integration constants we put $\bar{c}^A = 0$,
\bear{5.67}
&&c^A = \bar{\mu} (\bar{b}^A - b^A),
\\     \label{5.67a}
&&\bar{b}^A  = \bar{\mu} \sum_{r \in \bar{S}} \eta_r \nu_r^2  U^{rA}
                 - \bar{\mu} \delta^{A}_{1},
\\     \label{5.68}
&&C_0 = \bar{\mu}^2,
\\     \label{5.68a}
&&C_s =  \bar{\mu}^2 b_s^2, \qquad  b_s > 0,
\ear
where $\bar{\mu} > 0$,  $\bar{S} = \{ 0 \} \cup S$
and $\eta_0 \nu_0^2 = (U^0,U^0)^{-1}$.

The only essential restrictions imposed are the
inequalities  $C_0, C_s >0$ that cut a subclass in
the class of solutions from Section 4.  This
subclass contains non-extremal black hole solutions
and its "Kasner-like"  (non-black-hole) deformations.
For extremal black hole solutions  one should
consider the special case $C_0 = C_s =0$. (For
extremal black hole solutions  and its multicenter
generalizations see \cite{IMBl}.)

Due to  (\ref{5.42o}) the
parameters $b_s$, $s \in S$, are coinciding inside blocks:
\bear{5.42b}
b_s = b_{s'},
\ear
$s,s' \in S_i$, $i = 1, \ldots, k$.

It may be verified that the restrictions
(\ref{5.47n}) and (\ref{5.49n})  are satisfied
identically if and only if
\bear{5.68b}
U^0(b)= U^0_A b^A = -b^0+ \sum_{j=0}^n d_j b^j= 1,
\\ \label{5.68c}
U^s(b)= U^s_A b^A=
\sum_{i\in I_s}d_i b^i-\chi_s \lambda_{a_s \alpha} b^\alpha= 1,
\ear
$s\in S$.
This follows from identities $U^0(\bar{b})= 1$ and $U^s(\bar{b})= 1$,
$s \in S$.

Relation  (\ref{5.55n}) reads
\bear{5.68d}
\sum_{s \in S} \eta_s \nu_s^2 (b_s^2 - 1) +
h_{\alpha\beta} b^\alpha b^\beta +
\sum_{i=1}^n d_i (b^i)^2 +
\frac1{d_0-1} \left(\sum_{i=1}^n d_i b^i \right)^2 = \frac{d_0}{d_0-1},
\ear
where the relation (equivalent to (\ref{5.68b}) )
\bear{5.68e}
b^0 = \frac1{1 - d_0} \left[ \sum_{j=1}^n d_j b^j -1 \right],
\ear
is used.

Now we rewrite a solution (under restrictions imposed)
in a so-called "Kasner-like"  form that is more suitable for
analysing the behaviour at large distances
and for singling out the black hole solutions.
For this reason
we introduce a new radial variable $R = R(u)$ by relations
\beq{5.69}
\exp( - 2\bar{\mu} u) = 1 - \frac{2\mu}{R^{\bar{d}}},  \quad
\mu = \bar{\mu}/ \bar{d} >0, \quad  \bar{d} = d_0 -1,
\eeq
$u > 0$, $R^{\bar{d}} > 2\mu$.
For the function
\beq{5.69a}
f_s(u) = \frac{|Q_s|}{2\bar{\mu} b_s |\nu_s|}
[ \exp(\bar{\mu} b_s (u -u_s))
+ \eta_s \eps_s \exp(- \bar{\mu} b_s (u -u_s)) ]
\eeq
we put the restriction $f_s(0) = 1$, or, equivalently,
\beq{5.70}
\exp( -\bar{\mu} b_s u_s) +
\eta_s \eps_s \exp( \bar{\mu} b_s u_s) =
\frac{2\bar{\mu} b_s |\nu_s|}{|Q_s|}.
\eeq
This restriction guarantees
the  asymptotical flatness of the
 $(2+d_0)$-dimensional section of the metric
in the limit $R \to + \infty$ (or, when, $u \to + 0 $)).
It follows from (\ref{5.70}) that $u_s < 0$
for $\eta_s \eps_s = -1$.  In any case $f_s(u) > 0$
for $u \geq 0$.

Then,  solutions for the metric and scalar fields
(see (\ref{5.46n}), (\ref{5.63n})) may be written as follows
\bear{5.72n}
g= \Bigl(\prod_{s \in S}
\bar{H}_s^{2 \eta_s d(I_s)\nu_s^2/(D-2)} \Bigr)
\biggl\{ F^{b^0 -1} dR \otimes dR
+ R^2  F^{b^0} d \Omega^2_{d_0}  \\ \nn
-  \Bigl(\prod_{s \in S} \bar{H}_s^{-2 \eta_s \nu_s^2} \Bigr)
F^{b^1}  dt \otimes dt
+ \sum_{i = 2}^{n} \Bigl(\prod_{s\in S}
  \bar{H}_s^{-2 \eta_s \nu_s^2 \delta_{iI_s}} \Bigr)
  F^{b^i} g^i  \biggr\},
\\  \label{5.73}
\varphi^\alpha=
\sum_{s\in S} \eta_s \nu_s^2 \chi_s \lambda_{a_s}^\alpha
\ln \bar{H}_s + \frac{1}{2} b^{\alpha} \ln F,
\ear
where
\bear{5.73a}
F = 1 - \frac{2\mu}{R^{\bar d}},
\\ \label{5.73b}
\bar{H}_s = \hat{H}_s  F^{(1-b_s)/2},
\\ \label{5.74}
\hat{H}_s = 1 +  \hat{P}_s \frac{(1 - F^{b_s})}{2 \mu b_s },
\ear
\bear{5.74b}
\hat{P}_s = - \eps_s \eta_s P_s,
\\ \label{5.74a}
P_s = \frac{|Q_s|}{\bar{d} |\nu_s|} \exp( \mu u_s) > 0,
\ear
$s \in S$.
Due to  (\ref{5.42o})-(\ref{5.42s})
parameters $P_s$ and $\hat{P}_s$
are coinciding inside blocks:
\bear{5.42t}
P_s = P_{s'}, \qquad \hat{P}_s = \hat{P}_{s'},
\ear
$s,s' \in S_i$, $i = 1, \ldots, k$.
Parameters $b_s$ are
also coinciding inside blocks, see (\ref{5.42b}).
Parameters $b_s, b^i, b^{\alpha}$ obey the
relations (\ref{5.68c})-(\ref{5.68e}).

The fields of forms are given by  (\ref{2.28n}), (\ref{2.29n})
with
\bear{5.75}
&&\Phi^s = \frac{\nu_s}{H_s^{'}},
\\ \label{5.76}
&&H_s^{'}= \Bigl[1 - P_s^{'} \hat{H}_s^{-1}
\frac{(1 - F^{b_s})}{2 \mu b_s }  \Bigr]^{-1},
\\ \label{5.77}
&&P_s^{'} = - \frac{Q_s}{\nu_s \bar{d}}.
\ear
$s \in S$. It follows from (\ref{5.70}), (\ref{5.74}), (\ref{5.74b})
and (\ref{5.77})
that
\beq{5.78}
(P_s^{'})^2 =
P_s (\hat{P}_s +2 b_s \mu) =
- \eps_s \eta_s \hat{P}_s (\hat{P}_s +2 b_s \mu),
\eeq
$s \in S$. This relation is self-consistent,
i.e. its left- and right-hand sides have the same sign,
since due to (\ref{5.70}) and (\ref{5.74a})
\beq{5.79a}
P_s < 2 \mu b_s
\eeq
for $\eps_s \eta_s = +1$ and hence
\beq{5.79b}
\hat{P}_s > - 2 b_s \mu,
\eeq
for all $s \in S$.

\subsection{\bf Black hole  solutions}

Here we show that the  black hole solution
>from \cite{IMBl} may be obtained from
our spherically-symmetric  solutions
(\ref{5.72n})-(\ref{5.78})  when
\bear{5.or}
b^1 = b_s = 1, \qquad b^i =  b^{\alpha} = 0,
\ear
$s \in S$, $i =0,2, \ldots, n$, $\alpha = 1, \ldots, l$.

Under relations  (\ref{5.or}) imposed
the metric and scalar fields
(\ref{5.72n}) and (\ref{5.73})
read
\bear{5.72no}
g=
\Bigl(\prod_{s \in S} \hat{H}_s^{2 \eta_s d(I_s)\nu_s^2/(D-2)} \Bigr)
\biggl\{ \frac{dR \otimes dR}{1 - 2\mu / R^{\bar d}}
+ R^2  d \Omega^2_{d_0}  \\ \nn
- \Bigl(\prod_{s \in S} \hat{H}_s^{-2 \eta_s \nu_s^2} \Bigr)
 \left(1 - \frac{2\mu}{R^{\bar d }} \right)  dt \otimes dt
+ \sum_{i = 2}^{n} \Bigl(\prod_{s\in S }
  \hat{H}_s^{-2 \eta_s \nu_s^2 \delta_{iI_s}} \Bigr) g^i  \biggr\}, \\
\label{5.73n}
\varphi^\alpha=
\sum_{s\in S} \eta_s \nu_s^2 \chi_s \lambda_{a_s}^\alpha
\ln \hat{H}_s,
\ear
where $\mu > 0$, $R^{\bar{d}} > 2 \mu$ and
\beq{5.74n}
\hat{H}_s = 1 + \frac{\hat{P}_s}{R^{\bar{d}}}, \quad \hat{P}_s > - 2 \mu,
\eeq
$\hat{P}_s \neq 0$,  $s \in S$. Parameters $\hat{P}_s$ are coinciding
inside blocks (see (\ref{5.42t})).

The fields of forms are given by  (\ref{2.28n}), (\ref{2.29n})
with
\bear{5.75n}
&&\Phi^s = \frac{\nu_s}{H_s^{'} },  \\
\label{5.76n}
&&H_s^{'}= \Bigl(1 - \frac{P_s^{'}}{\hat{H}_s R^{\bar{d}}} \Bigr)^{-1} =
1 + \frac{P_s^{'}}{ R^{\bar{d}} + \hat{P}_s - P_s^{'} },
\ear
$s \in S$. Here
\beq{5.78n}
(P_s^{'})^2 =  - \eps_s \eta_s \hat{P}_s (\hat{P}_s +2\mu),
\eeq
and
\bear{5.78m}
\eps_s \eta_s \hat{P}_s  < 0,
\ear
$s \in S$. Parameters $\nu_s$ satisfy relations
(\ref{5.42p}).

The solution obtained describes non-extremal charged
$p$-brane black holes with
block-orthogonal intersection rules. The exteriour
horizon corresponds to $R^{\bar{d}} \to 2\mu$.

Let
\bear{5.sr}
\eps_s \eta_s = -1,
\ear
$s \in S$. This restriction is satisfied in orthogonal case,
when pseudo-Euclidean $p$-branes
in a space-time of  pseudo-Euclidean signature are
considered (in this case all $\eps(I_s) = -1$,
$\eps[g] = -1$), all $\theta_s =
+1$ in the action (\ref{2.1i}) and $\eta_s= {\rm sign}(U^s,U^s) = +1$).

Under restrictions (\ref{5.sr}) imposed
our solutions agree with
the solutions  with  orthogonal intersection rules
>from Refs. \cite{CT}, \cite{AIV}, \cite{Oh}
($d_1 = \ldots = d_n =1$, $\eta_s = + 1$), \cite{BIM}
($\eta_s = + 1$, non-composite case)
and block-orthogonal ones from
\cite{Br1} (for $\nu_s$ coinciding inside blocks).

{\bf Hawking temperature.}
The Hawking temperature corresponding to
the solution is (see also \cite{BIM,Oh})
\beq{5.79n}
T_H(\mu) =
\frac{\bar{d}}{4 \pi (2 \mu)^{1/\bar{d}}}
\prod_{s \in S}
\left(\frac{2 \mu}{2 \mu + \hat{P}_s}\right)^{\eta_s \nu_s^2}.
\eeq

For fixed $\hat{P}_s > 0$  ($\eps_s \eta_s = -1$)
and $\mu \to + 0$ we get
$T_H(\mu) \to 0$ for the extremal
black hole configurations \cite{IMBl} satisfying
\beq{5.80}
\xi = \sum_{s\in S}\eta_s \nu_s^2- \bar{d}^{-1} > 0.
\eeq

\section{Post-Newtonian approximation}

Let $d_0 = 2$. Here we consider the 4-dimensional section of the
metric (\ref{5.72n})
\bear{4.1}
g^{(4)} = U \biggl\{ F^{b^0-1} dR \otimes dR
+ F^{b^0} R^2  d \hat{\Omega}^2_{2}
- U_1 F^{b^1}   dt \otimes dt \biggr\},
\ear
where  $F = 1- (2 \mu/R)$, and
\bear{4.1a}
&&U = \prod_{s \in S} \bar{H}_s^{2 \eta_s d(I_s) \nu_s^2/(D-2)},
\\ \label{4.2a}
&&U_1 = \prod_{s \in S} \bar{H}_s^{-2 \eta_s \nu_s^2},
\ear
$R > 2\mu$.

We may suppose that some real astrophysical objects (e.g. stars)
are described by the 4-dimensional "physical" metric (\ref{4.1}),
i.e. they are "traces" of extended multidimensional objects
(charged $p$-branes).

Introducing a new radial variable $\rho$ by the relation
\bear{4.2}
R = \rho \left(1 + \frac{\mu}{2\rho}\right)^2,
\ear
($\rho > \mu/2$), we rewrite the metric (\ref{4.1})
in a 3-dimensional conformally-flat form
\bear{4.3}
g^{(4)} = U \Biggl\{ - U_1 F^{b^1} dt \otimes dt +
F^{b^0} \left(1 + \frac{\mu}{2 \rho} \right)^4
\delta_{ij} dx^i \otimes dx^j \Biggr\},
\\ \label{4.3a}
F =  \left(1 - \frac{\mu}{2 \rho} \right)^2
\left(1 + \frac{\mu}{2 \rho} \right)^{-2}
\ear
where $\rho^2 =|x|^2 = \delta_{ij}x^i x^j$ ($i,j = 1,2,3$).

For possible physical applications we should calculate the
post-Newtonian parameters $\beta$ and $\gamma$ (Eddington parameters)
using the following relations (see, for example, \cite{Dam} and
references therein)
\bear{4.4}
g^{(4)}_{00} = - (1 -  2 V + 2 \beta V^2 ) + O(V^3),
\\
\label{4.5}
g^{(4)}_{ij} = \delta_{ij}(1 + 2 \gamma V ) + O(V^2),
\ear
$i,j = 1,2,3$, where
\bear{4.6}
V = \frac{GM}{\rho}
\ear
is the Newton's potential, $G$ is the gravitational constant,
$M$ is the gravitational mass.
>From (\ref{4.3})-(\ref{4.6}) we get
\bear{4.7}
GM = \mu b^1 + \sum_{s \in S} \eta_s \nu_s^2
[ \hat{P}_s + (b_s - 1)\mu ]
\left(1 -  \frac{d(I_s)}{D-2} \right)
\ear
and  for $GM \neq 0$
\bear{4.8}
\beta - 1 = \frac{1}{2(GM)^2} \sum_{s \in S}  \eta_s \nu_s^2
\hat{P}_s (\hat{P}_s + 2 b_s \mu)
\left(1 -  \frac{d(I_s)}{D-2} \right)
\\ \label{4.9}
\gamma - 1 = - \frac{1}{GM}
\left[ \mu (b^0 + b^1 - 1)  +  \sum_{s \in S} \eta_s \nu_s^2
[\hat{P}_s + (b_s -1)\mu ]
\left(1 -  2 \frac{d(I_s)}{D-2} \right) \right].
\ear

It follows from  (\ref{5.78}), (\ref{4.8})
and the inequalities  $d(I_s) < D - 2$ (for all $s \in S$)
that the following inequalities take place
\bear{4.10}
\beta > 1, {\rm \ if \  all} \ \eps_s = -1,
\\ \label{4.10a}
\beta < 1, {\rm \ if \  all} \ \eps_s = +1.
\ear
There exists a large variety of configurations
with $\beta = 1$ when the relation $\eps_s = {\rm const}$
is broken.

There exist also non-trivial $p$-brane configurations with $\gamma =1$.

{\bf Proposition.}
Let the set of $p$-branes consist of
several pairs of electric and magnetic branes. Let
any such pair $(s, \bar{s} \in S)$  correspond to the same colour index,
i.e. $a_s = a_{\bar{s}}$, and $\hat{P}_s = \hat{P}_{\bar{s}}$,
$b_s = b_{\bar{s}}$, $\eta_s \nu_s^2 = \eta_{\bar{s}} \nu_{\bar{s}}^2$.
Then  for $b^0 + b^1 = 1$  we get
\bear{4.13}
\gamma = 1.
\ear

The Proposition can be readily proved using the relation
$d(I_s) + d(I_{\bar{s}}) = D - 2$, following from (\ref{2.d1})
and (\ref{2.d2}).

{\bf Observational restrictions.} The observations in the solar system
give the tight constraints on the Eddington parameters \cite{Dam}
\bear{4.14}
\gamma = 1.000 \pm 0.002
\\
\label{4.15}
\beta = 0.9998 \pm 0.0006.
\ear
The first restriction is a result of the Viking time-delay experiment
\cite{Re}. The second restriction follows from (\ref{4.14})
and the analysis of the laser ranging data to the Moon.
In this case a high precision test based on the calculation
of the combination
$(4\beta - \gamma - 3)$ appearing in the Nordtvedt effect
\cite{N} is used \cite{Di}.
We note, that as it was pointed in \cite{Dam}
the "classic" tests of general relativity, i.e. the
Mercury-perihelion and light deflection tests, are somewhat
outdated.

For small enough
$\hat{p}_s = \hat{P}_s/GM$, $b_s - 1$, $b^1 - 1$, $b^i$ ($i > 1$)
of the same order we get $GM \sim \mu$ and hence
\bear{4.17}
\beta - 1 \sim  \sum_{s \in S} \eta_s \nu_s^2 \hat{p}_s
\left(1 -   \frac{d(I_s)}{D-2} \right)
\\ \label{4.17a}
\gamma - 1 \sim -b^0 - b^1 + 1
- \sum_{s \in S} \eta_s \nu_s^2 [\hat{p}_s + (b_s -1)]
\left(1 -  2 \frac{d(I_s)}{D-2} \right) ,
\ear
i.e. $\beta -1$ and $\gamma - 1$ are of the same order.
Thus for small enough
$\hat{p}_s $, $b_s - 1$, $b^1 - 1$, $b^i$ ($i > 1$)
it is possible to fit
the "solar system" restrictions (\ref{4.14}) and (\ref{4.15}).

There exists also
another possibility to satisfy these restrictions.

{\bf One brane case.}
Let us consider a special  case of one $p$-brane.
In this case we have
\bear{4.18}
\eta_s \nu^{-2}_s =
d(I_s) \left(1 -  \frac{d(I_s)}{D-2} \right) + \lambda^2.
\ear
Relations (\ref{4.8}),  (\ref{4.9})  and   (\ref{4.18})
imply that for large enough values of
(dilatonic coupling constant squared)  $\lambda^2$
and $b^0 + b^1 = 1$ it is possible to perform
the "fine tuning"  the parameters $(\beta, \gamma)$ near
the point (1, 1) even if the parameters  $\hat{P}_s$ are big.

\section{Conclusions}

In this paper we obtained exact solutions to Einstein
equations for the multidimensional cosmological
model describing the evolution of $n$ Ricci-flat spaces
and one Einstein space $M_0$  of non-zero curvature
in the presence of composite electro-magnetic $p$-branes.
The solutions were obtained in the block-orthogonal case (\ref{5.4n}),
when $p$-branes do not "live" in $M_0$. We also considered the
spherically-symmetric solutions containing non-extremal
$p$-brane black holes \cite{Br1,IMBl}. The relations
for post-Newtonian parameters $\beta$ and $\gamma$ are obtained.

\bigskip

\noi
{\bf Acknowledgments.}
This work was supported in part
by DFG grant 436 RUS 113/236/O(R) and
by the Russian Ministry for
Science and Technology,  Russian Fund for Basic Research,
project N 98-02-16414 and project SEE.

\renewcommand{\thesection}{}
\section{Appendix 1: Restrictions on $p$-brane configurations}

{\bf Restrictions on $\Omega_{a,v}$ \cite{IMJ}.} Let
\beq{B.12i}
w_1\equiv\{i \mid i\in\{0,\dots,n\},\ d_i=1\}.
\eeq
The set $w_1$ describes all $1$-dimensional manifolds among $M_i$ $(i\ge0)$.
We impose the following restrictions on the sets $\Omega_{a,v}$
(\ref{2.29nn}):
\beq{B.31in}
W_{ij}(\Omega_{a,v})=\emptyset,
\eeq
$a\in\tri$; $v=e,m$; $i,j\in w_1$, $i<j$ and
\beq{B.17in}
W_j^{(1)}(\Omega_{a,m},\Omega_{a,e})=\emptyset,
\eeq
$a\in\tri$; $j\in w_1$. Here
\bear{B.23i}
W_{ij}(\Omega_*)\equiv
\{(I,J)|I,J\in\Omega_*,\ I=\{i\}\sqcup(I\cap J),\
J=\{j\}\sqcup(I\cap J)\},
\ear
$i,j\in w_1$, $i\ne j$, $\Omega_* \subset \Omega_0$ and
\beq{B.17i}
W_j^{(1)}(\Omega_{a,m},\Omega_{a,e}) \equiv
\{(I,J)\in\Omega_{a,m}\times\Omega_{a,e}|\bar I=\{j\}\sqcup J\},
\eeq
$j\in w_1$. In (\ref{B.17i})
\beq{B.28i}
\bar I\equiv I_0 \setminus I
\eeq
is "dual" set, ($I_0 = \{0,1,\ldots,n \}$).

The restrictions (\ref{B.31in}) and (\ref{B.17in}) are
trivially satisfied when $n_1\le1$ and $n_1=0$ respectively, where
$n_1=|w_1|$ is the number of $1$-dimensional manifolds among $M_i$.
They are also satisfied in the non-composite case
when all $|\Omega_{a,v}| = 1$.
For $n_1\ge 2$  and $n_1 \ge 1$, respectively,
these restrictions  forbid certain
pairs of two $p$-branes,
corresponding to the same form $F^a, a \in \tri$:

\section{Appendix 2: Solutions with \protect\\
block-orthogonal set of vectors }

Let
\bear{A.1}
L=\frac12<\dot x,\dot x>- \sum_{s \in S} A_s\exp(2<b_s,x>)
\ear
be a Lagrangian, defined on $V\times V$, where $V$ is
a $n$-dimensional vector space over $\R$, $A_s\ne0$, $s \in S$;
$S \ne\emptyset$, and $<\cdot,\cdot>$ is a
non-degenerate real-valued quadratic form on $V$.
Let
\bear{A.2a}
S=S_1\sqcup\dots\sqcup S_k,
\ear
all $S_i\ne\emptyset$, and
\bear{A.2b}
<b_s,b_{s'}>=0,
\ear
for all $s \in S_i$, $s' \in S_j$,
$i\ne j$; $i,j=1,\dots,k$.

Let us suppose that there exists a set $h_s \in \R$, $h_s \neq 0$,
$s \in S$, such that
\bear{A.2}
\sum_{s \in S} <b_s,b_{s'}> h_{s'} = -1,
\ear
for all $s \in S$, and
\bear{A.3}
\frac{A_s}{h_s} = \frac{A_{s'}}{h_{s'}},
\ear
$s,s' \in S_i$, $i = 1, \ldots, k$,
(the ratio $A_s/h_s$ is constant inside  $S_i$).

Then, the Euler-Lagrange equations for the Lagrangian (\ref{A.1})
\beq{A.4}
\ddot{x} + \sum_{s \in S} 2 A_s b_s \exp(2<b_s,x>) =0,
\eeq
have the following special solutions
\beq{A.5}
x(t)= \frac12 \sum_{s \in S} h_s b_s
\ln \left[y_s^2(t) \left|\frac{2A_s}{h_s}\right|\right] + \alpha t + \beta,
\eeq
where $\alpha,\beta\in V$,
\beq{A.6}
<\alpha,b_s>=<\beta,b_s>=0,
\eeq
$s \in S$, and functions $y_s(t) \neq 0$
satisfy the equations
\beq{A.7}
\frac{d}{dt}\left(y_s^{-1} \frac{d y_s}{dt}\right) = - \xi_s y_s^{-2},
\eeq
with
\beq{A.8}
\xi_s = {\rm sign} \left(\frac{A_s}{h_s}\right),
\eeq
$s \in S$, and coincide inside blocks:
\bear{A.9}
y_s(t) = y_{s'}(t),
\ear
$s,s' \in S_i$, $i = 1, \ldots, k$.
More explicitly
\bear{A.10}
y_s(t) = s(t - t_s, \xi_s, C_s),
\ear
where constants $t_s, C_s \in \R$ coincide inside blocks
\bear{A.11}
t_s = t_{s'}, \qquad C_s = C_{s'},
\ear
$s,s' \in S_i$, $i = 1, \ldots, k$,
and
\bear{A.12}
s(t, \xi, C) \equiv
\frac{1}{\sqrt{C}} \sinh(t \sqrt{C}), \ \xi = +1, \quad C>0; \\
\label{A.13}
\frac{1}{\sqrt{-C}} \sin(t \sqrt{-C}), \ \xi = +1, \quad C <0; \\
\label{A.14}
t, \ \xi = +1, \quad C = 0; \\
\label{A.15}
\frac{1}{\sqrt{C}} \cosh(t \sqrt{C}), \ \xi = -1, \quad C>0.
\ear

For the energy
\bear{A.16}
E=\frac12<\dot x,\dot x> + \sum_{s \in S} A_s\exp(2<b_s,x>)
\ear
corresponding to the solution (\ref{A.5}) we have
\beq{A.17}
E= \frac12 \sum_{s \in S} C_s (- h_s) + \frac12 <\alpha,\alpha> .
\eeq

For dual vectors $u^s\in V^*$ defined as
$u^s(x)=<b_s,x>$, $\forall x \in V$, we have $<u^s,u^l>_*=<b_s,b_l>$,
where $< \cdot, \cdot>_*$ is dual form on  $V^*$.  The orthogonality
conditions (\ref{A.6}) read
\beq{A.18}
u^s(\alpha)=u^s(\beta)=0 ,
\eeq
$s \in S$.


\small
\begin{thebibliography}{199}

\bibitem{Mel2}
V.N. Melnikov, Multidimensional Classical and Quantum Cosmology and
Gravitation.Exact Solutions and Variations of Constants. CBPF-NF-051/93,
Rio de Janeiro, 1993; \\
V.N. Melnikov. In: {\it Cosmology and Gravitation}, ed. M.Novello
(Editions Frontieres, Singapore, 1994) p. 147.
\bibitem{Mel}
V.N. Melnikov,
Multidimensional Cosmology and  Gravitation,
CBPF-MO-002/95, Rio de Janeiro, 1995, 210 p. \\
V.N. Melnikov. In {\it Cosmology and Gravitation.II} ed. M. Novello
(Editions Frontieres, Singapore, 1996) p. 465.
\bibitem{3}
K.P. Staniukovich and V.N. Melnikov, {\it Hydrodynamics, Fields and
Constants
in the Theory of Gravitation}, (Energoatomizdat, Moscow, 1983), (in
Russian).

\bibitem{HTW}
C. Hull and P. Townsend, Unity of Superstring Dualities,
{\it Nucl. Phys.\/} {\bf B 438}, 109 (1995). \\
P. Horava and E. Witten, {\it Nucl. Phys.\/} {\bf B 460}, 506 (1996).
\bibitem{Hull}
C.M. Hull, String dynamics at strong coupling,
{\it Nucl. Phys.} {\bf B 468}, 113 (1996).
\bibitem{Sc}
J.M. Schwarz,  Lectures on Superstring and M-Theory Dualities,
hep-th/9607201;
\bibitem{Du}
M.J. Duff,  M-theory (the Theory Formerly Known as Strings),
hep-th/9608117.
\bibitem{Vafa}
C. Vafa, Evidence for F-Theory, hep-th/9602022;
{\it Nucl. Phys.} {\bf B 469}, 403 (1996).
\bibitem{Nic2}
H. Nicolai, On M-theory, hep-th/9801090.
\bibitem{4}
V.N. Melnikov, {\it Int. J. Theor.Phys.} {\bf 33}, N7, 1569 (1994).
\bibitem{5}
V. de Sabbata, V.N.Melnikov and P.I.Pronin,
{\it Prog. Theor. Phys.} {\bf 88}, 623 (1992).
\bibitem{6}
V.N. Melnikov. In: {\it Gravitational Measurements, Fundamental Metrology
and Constants}. Eds. V. de Sabbata and V.N. Melnikov (Kluwer Academic Publ.)
Dordtrecht, 1988, p.283.
\bibitem{7}
A.J. Sanders and G.T. Gillies, {\it Rivista Nuovo Cim.} {\bf 19},
N2, 1 (1996).
\bibitem{8}
A.J. Sanders and G.T. Gillies, {\it Grav. and Cosm.} {\bf 3}, N4(12),
285 (1997).
\bibitem{SEE}
A.J. Sanders and W.E. Deeds. {\it Phys. Rev. } {\bf D 46}, 480 (1992).

\bibitem{9}
G.T. Gillies, {\it Rep. Progr. Phys.} {\bf 60}, 151 (1997).
\bibitem{10}
V. Achilli et al., {\it Nuovo Cim.} {\bf B 12}, 775 (1997).
\bibitem{Kal}
T. Kaluza, {\it Sitzungsber. Preuss. Akad. Wiss. Berlin
Phys. Math.}, {\bf K1} 33, 966 (1921).
\bibitem{Kl}
O. Klein, {\it Z. Phys.} {\bf 37}, 895 (1926).
\bibitem{DeS}
V. De Sabbata and E. Schmutzer, {\it Unified Field Theories in more
than Four Dimensions}, (World Scientific, Singapore, 1982).
\bibitem{Lee}
H. C. Lee, {\it An Introduction to Kaluza-Klein Theories},
(World Scientific, Singapore, 1984).
\bibitem{Vl1}
Yu.S. Vladimirov {\it Physical Space-Time Dimension and Unification of
Interactions} (University Press, Moscow, 1987) (in Russian).
\bibitem{WeP1}
P.S. Wesson and J. Ponce de Leon,
{\it Gen. Rel. Gravit.} {\bf 26}, 555 (1994).
\bibitem{J}
P. Jordan, Erweiterung der projektiven Relativitatstheorie,
{\it Ann. der Phys.}  219 (1947).
\bibitem{BD}
C. Brans and R.H. Dicke, {\it Phys. Rev. D} {\bf 124}, 925 (1961).
\bibitem{CJS}
E. Cremmer, B. Julia, and J. Scherk, {\it Phys. Lett.} {\bf B76}
409 (1978).
\bibitem{SaSe}
A. Salam and E. Sezgin, eds.,
{\it Supergravities in Diverse Dimensions}, reprints in 2 vols.,
(World Scientific, Singapore, 1989).
\bibitem{GrSW}
M.B. Green, J.H. Schwarz and E. Witten, {\it Superstring Theory}
(Cambridge University Press., Cambridge, 1987).
\bibitem{BelKh}
V.A. Belinskii and I.M. Khalatnikov, {\it ZhETF}, {\bf 63}, 1121 (1972).
\bibitem{FH}
P. Forgacs and Z. Horvath, {\it Gen. Rel. Grav.} {\bf 11}, 205 (1979).
\bibitem{ChD}
A. Chodos and S. Detweyler, {\it Phys. Rev.} {\bf D 21}, 2167 (1980).
\bibitem{Fr}
P.G.O. Freund, {\it Nucl. Phys. } {\bf B 209}, 146 (1982).
\bibitem{ABE}
R. Abbot, S. Barr and S. Ellis, {\it Phys. Rev.} {\bf D 30}, 720 (1984).
\bibitem{RubS}
V.A. Rubakov and M.E. Shaposhnikov, {\it Phys. Lett. } {\bf B 125}
136 (1983).
\bibitem{Sah}
D. Sahdev, {\it Phys. Lett.} {\bf B 137}, 155 (1984).
\bibitem{KoLS}
E. Kolb, D. Linkley and D. Seckel, {\it Phys. Rev. } {\bf D 30} 1205 (1984).
\bibitem{RSS}
S. Ranjbar-Daemi, A. Salam and J. Strathdee, {\it Phys. Lett.}
{\bf B 135}, 388 (1984).
\bibitem{Lor1}
D. Lorentz-Petzold, {\it Phys. Lett. } {\bf B 148} 43 (1984).
\bibitem{BO}
R. Bergamini and C.A. Orzalesi, {\it Phys. Lett. } {\bf B 135}, 38 (1984).
\bibitem{GRT}
M. Gleiser, S. Rajpoot and J.G. Taylor, {\it Ann.
Phys. (NY)} {\bf 160}, 299 (1985).
\bibitem{BL1}
U. Bleyer and D.-E. Liebscher, in
{\it Proc. III Sem. Quantum Gravity}  ed. M.A. Markov, V.A. Berezin and
V.P. Frolov  (Singapore, World Scientific, 1985) p. 662.
\bibitem{BL2}
U. Bleyer and D.-E. Liebscher, {\it Gen. Rel. Gravit.} {\bf 17},
989 (1985).
\bibitem{DeGHS}
M. Demianski, Z. Golda, M. Heller and M. Szydlowski,
{\it Class. Quantum Grav.} {\bf 3}, 1190 (1986).
\bibitem{Wil}
D.L. Wiltshire, {\it Phys. Rev. } {\bf D 36}, 1634 (1987).
\bibitem{BL3}
U. Bleyer and D.-E. Liebscher, {\it Annalen d. Physik (Lpz)} {\bf 44}
81 (1987).
\bibitem{GibM}
G.W. Gibbons and K. Maeda, {\it Nucl. Phys. B} {\bf 298}, 741 (1988).
\bibitem{WW}
Y.-S. Wu and Z. Wang, {\it Phys. Rev. Lett.} {\bf 57} 1978 (1986).
\bibitem{GibW}
G.W. Gibbons and D.L. Wiltshire, {\it Nucl. Phys.} {\bf B 287},
717 (1987).
\bibitem{IM1}
V.D. Ivashchuk and V.N. Melnikov, {\it Nuovo Cimento} {\bf B 102},
131 (1988).
\bibitem{BIM1}
K.A. Bronnikov, V.D. Ivashchuk and V.N. Melnikov,
{\it Nuovo  Cimento} {\bf B 102}, 209 (1988).
\bibitem{IM2}
V.D. Ivashchuk and  V.N. Melnikov, {\it Phys. Lett. } {\bf A 135},
465 (1989).
\bibitem{IMZ}
V.D. Ivashchuk, V.N. Melnikov and A.I. Zhuk,
{\it Nuovo Cimento } {\bf B 104}, 575  (1989).
\bibitem{Ber}
V.A. Berezin, G. Domenech, M.L. Levinas, C.O. Lousto and N.D. Umerez,
{\it Gen. Relativ. Grav.} {\bf 21}, 1177 (1989).
\bibitem{IM3A}
V.D. Ivashchuk and V.N. Melnikov,
{\it Chinese Phys. Lett.} {\bf 7}, 97 (1990).
\bibitem{DP}
M. Demiansky and A. Polnarev, {\it Phys. Rev. } {\bf D 41}, 3003 (1990).
\bibitem{FIM1}
S.B. Fadeev, V.D. Ivashchuk and V.N. Melnikov in {\it Gravitation
and Modern Cosmology} (Plenum, N.-Y., 1991) p. 37.
\bibitem{BLP}
U. Bleyer, D.-E. Liebscher and A.G. Polnarev, {\it  Class. Quant. Grav.}
{\bf 8}, 477 (1991).
\bibitem{I1}
V.D. Ivashchuk, {\it Phys. Lett. } {\bf A 170}, 16 (1992).
\bibitem{Zh1}
A. Zhuk, {\it Class. Quant. Grav.} {\bf 9},  202 (1992).
\bibitem{Zh2}
A. Zhuk, {\it Phys.  Rev.} {\bf D 45},  1192 (1992).
\bibitem{Zh3}
A. Zhuk, {\it Sov. Journ. Nucl. Phys.} {\bf 55}, 149 (1992).

\bibitem{Zh4}
A.I. Zhuk, {\it Sov. Journ. Nucl. Phys.} {\bf 56}, 223  (1993).

\bibitem{Mis}
C.W. Misner, In: Magic without Magic: John Archibald Wheeler, ed.
J.R. Klauder, Freeman, San Francisko, 1972.

\bibitem{Hal}
J.J. Halliwell, {\it Phys. Rev.} {\bf D 38}, 2468  (1988).
\bibitem{HawP}
S.W. Hawking and D.N. Page, {\it Phys. Rev. D} {\bf 42}, 2655 (1990).
\bibitem{LiWP}
H. Liu, P.S. Wesson and J. Ponce de Leon, {\it J. Math. Phys.}
{\bf 34}, 4070 (1993).
\bibitem{Gav}
V.R. Gavrilov,  {\it Hadronic J.} {\bf 16}, 469 (1993).

\bibitem{IMT}
V.D. Ivashchuk and V.N. Melnikov, {\it Teor.  Mat. Fiz.}
{\bf 98}, 312 (1994) (in Russian).

\bibitem{IM3}
V.D. Ivashchuk and V.N. Melnikov,
Multidimensional cosmology with $m$-component perfect fluid,
gr-qc/ 9403063;
{\it Int. J. Mod. Phys.} {\bf D 3}, 795 (1994).

\bibitem{GIM}
V.R. Gavrilov, V.D. Ivashchuk and  V.N. Melnikov,
Integrable pseudo-euclidean Toda-like systems in multidimensional
cosmology with multicomponent perfect fluid, {\it J. Math. Phys }
{\bf 36}, 5829 (1995).

\bibitem{BZ1}
U.Bleyer and A. Zhuk, On multidimensional cosmological models with static
internal spaces, {\it Class. and Quantum Grav.}
{\bf 12}, 89 (1995).

\bibitem{BZ2}
U. Bleyer and A. Zhuk,
Multidimensional integrable cosmological models with negative external
curvature, {\it  Grav. and Cosmol.}, {\bf 2} 106 (1995).

\bibitem{BZ2A}
U. Bleyer and A. Zhuk, Multidimensional integrable cosmological models
with positive external space curvature,
{\it Grav. and Cosmol.} {\bf 1}, 37 (1995).

\bibitem{BZ3}
U. Bleyer and A. Zhuk, Kasner-like, inflationary and steady-state solutions
in multidimensional cosmology, { \it Astron. Nachrichten} {\bf 317},
161 (1996).

\bibitem{Zh5}
A.I. Zhuk, {\it Sov. Journ. Nucl. Phys.} {\bf 58},  No 11 (1995).


\bibitem{BS}
J.D. Barrow and J. Stein-Schabes, {\it Phys. Rev.} {\bf D 32}, 1595 (1985).
\bibitem{DHS}
J. Demaret, M. Henneaux and P. Spindel, {\it Phys. Lett.} {\bf B 164}
27 (1985).
\bibitem{SzP}
M. Szydlowski and G. Pajdosz, {\it Class. Quant. Grav.} {\bf 6} (1989),
1391.

\bibitem{IvKM1}
V.D. Ivashchuk, A.A. Kirillov and V.N. Melnikov,
{\it Izv. Vuzov (Fizika)}, {\bf 37}, No 11 (1994) 107 (in Russian).
\bibitem{IvKM}
V.D. Ivashchuk, A.A. Kirillov and V.N. Melnikov, {\it Pis'ma ZhETF } {\bf
60}
No 4 (1994), 225 (in Russian).
\bibitem{IM6}
V.D. Ivashchuk and V.N. Melnikov, Billiard
representation for multidimensional cosmology with multicomponent perfect
fluid near the singularity, {\it Class. Quantum Grav.}
{\bf 12}, 809 (1995).
\bibitem{KM}
A.A. Kirillov and V.N. Melnikov,
Dynamics Inhomogeneouties of Metric in the Vicinity of a Singularity
in Multidimensional Cosmology,
{\it Phys. Rev. } {\bf D 52}, 723 (1995). \\
A.A. Kirillov and V.N. Melnikov,
On Properties of Metrics Inhomogeneouties in the Vicinity of a Singularity
in K-K Cosmological Models, {\it Astron. Astrophys. Trans.}, {\bf 10},
101 (1996).
\bibitem{Rain}
M. Rainer, {\it Grav. and Cosmol. } {\bf 1}, 81 (1995).
\bibitem{GasVen1}
M. Gasperini and G. Veneziano,
{\it Phys. Rev. } {\bf D 50}, 2519 (1994).
\bibitem{GasVen2}
M. Gasperini and G. Veneziano,
{\it Mod. Phys. Lett.} {\bf A 8}, 701 (1993).
\bibitem{Ang}
C. Angelantonj, L. Amendola, M. Litterio and F. Occhionero,
String cosmology and inflation, {\it Phys. Rev. } {\bf D 51}, 1607 (1995).
\bibitem{Kr}
D. Kramer, {\it Acta Physica Polonica} {\bf 2}, F. 6,  807 (1969).
\bibitem{Le}
A.I. Legkii, in {\it Probl. of Grav. Theory and Elem. Particles}
(Atomizdat, Moscow)  {\bf 10}, 149 (1979) (in Russian).
\bibitem{GP}
D.J. Gross and M.J. Perry, {\it Nucl. Phys.} {\bf B 226}, 29 (1993).
\bibitem{Tan}
F.R. Tangherlini, {\it Nuovo Cimento} {\bf 27}, 636 (1963).
\bibitem{BrI}
K.A. Bronnikov, V.D. Ivashchuk in {\it Abstr.  VIII Soviet Grav. Conf}
(Erevan, EGU, 1988) p. 156.
\bibitem{FIM2}
S.B. Fadeev, V.D. Ivashchuk and V.N. Melnikov, {\it Phys. Lett.}
{\bf A 161}, 98 (1991).
\bibitem{FIM3}
S.B. Fadeev, V.D. Ivashchuk and V.N. Melnikov, {\it Chinese Phys. Lett.}
{\bf 8}, 439 (1991).
\bibitem{BM}
K.A. Bronnikov and V.N. Melnikov, {\it Annals of Physics (N.Y.)} {\bf 239},
40 (1995).
\bibitem{BI}
U. Bleyer and V.D. Ivashchuk, {\it Phys. Lett.} {\bf B 332}, 292 (1994).

\bibitem{IM8}
V.D. Ivashchuk and V.N. Melnikov,
{\it Class. Quantum Grav.}, {\bf 11}, 1793 (1994).

\bibitem{IM10}
V.D. Ivashchuk and V.N. Melnikov,
{\it Grav. and Cosmol.} {\bf 1}, No 3, 204 (1995).

\bibitem{IM13}
V.D. Ivashchuk and V.N. Melnikov,
Extremal Dilatonic Black Holes in String-like Model with Cosmological
Term, {\it Phys. Lett.} {\bf B 384}, 58 (1996).

\bibitem{Rub}
V.A. Rubakov, {\it Phys. Lett.} , {\bf B 214}, 503 (1988).
\bibitem{GidS2}
S. Giddings and A. Strominger, {\it Nucl. Phys. } {\bf B 321}, 481 (1989).
\bibitem{Kir}
A.A. Kirillov, {\it Pis'ma ZhETF}  {\bf 55}, 541 (1992).
\bibitem{GKag}
E.I. Guendelman and A.B. Kaganovich,
{\it Phys. Lett.} {\bf B 301}, 15 (1993).
\bibitem{Hor}
T. Horigushi, {\it Mod. Phys. Lett. } {\bf A 8},  777 (1993).

\bibitem{BIMZ}
U. Bleyer, V.D. Ivashchuk, V.N. Melnikov and A.I. Zhuk,
Multidimensional
classical and quantum wormholes in models with cosmological constant.
gr-qc/9405020; {\it Nucl. Phys.} {\bf B 429} 117  (1994).

\bibitem{GKMR}
V.R. Gavrilov, U. Kasper, V.N. Melnikov and M. Rainer,
Toda Chains with Type $A_m$ Lie Algebra for Multidimensional m-component
Perfect Fluid Cosmology, Preprint Math-97/ Univ. Potsdam, 1997.

\bibitem{GIM2}
V.R. Gavrilov, V.D. Ivashchuk, and V.N. Melnikov,
{\it  Class. Quant. Grav.} {\bf 13}, 3039 (1996).
\bibitem{GI}
V.R. Gavrilov and V.N. Melnikov,
{\it  Theor. Math. Phys} {\bf 114}, N3, 454 (1998).
\bibitem{GMT}
V.R. Gavrilov, V.N. Melnikov and R. Triay,
Exact Solutions in Multidimensional Cosmology with Shear and Bulk
Viscosity, {\it  Class. Quant. Grav.} {\bf 14}, 2203 (1997). \\
V.R. Gavrilov, V.N. Melnikov and M. Novello,
Exact Solutions in Multidimensional Cosmology with Bulk Viscosity,
{\it Grav. and Cosmol.} {\bf 1}, No 2, 149 (1995). \\
V.R. Gavrilov, V.N. Melnikov and M. Novello,
Bulk Viscosity and Entropy Production in Multidimensional Integrable
Cosmology {\it Grav. and Cosmol.} {\bf 2}, No 4(8), 325 (1996).
\bibitem{RZ}
M. Rainer and A. Zhuk, {\it Phys. Rev.}, {\bf D 54} 6186 (1996).

\bibitem{IME}
V.D. Ivashchuk and V.N. Melnikov,
Multidimensional Gravity with Einstein Internal spaces,
hep-th/9612054;  {\it Grav. and Cosmol.}
{\bf 2}, No 3 (7), 177 (1996).

\bibitem{BrF}
K.A. Bronnikov and J.C. Fabris,
{\it Grav. and Cosmol.} {\bf 2}, No 4 (8), (1996).

\bibitem{DKL}
M.J. Duff, R.R. Khuri and J.X. Lu,
{\it Phys. Rep.\/} {\bf 259},  213 (1995).
\bibitem{St}
K.S. Stelle, Lectures on Supergravity p-Branes, hep-th/9701088.
hep-th/9608117.
\bibitem{GHT}
G.W. Gibbons, G.T. Horowitz and P.K. Townsend,
{\it Class. Quant. Grav.} {\bf 12}, 297 (1995); hep-th/9410073.

\bibitem{Dab}
A. Dabholkar, G. Gibbons, J.A. Harvey, and F. Ruiz Ruiz,
{\it Nucl. Phys.} {\bf B 340}, 33 (1990).
\bibitem{CHS}
C.G. Callan, J.A. Harvey and A. Strominger,
{\it Nucl. Phys.} {\bf 359} (1991) 611;  {\it Nucl. Phys.}
{\bf B 367}, 60 (1991).
\bibitem{DS}
M.J. Duff and K.S. Stelle, {\it Phys. Lett.} {\bf B 253},  113 (1991).
\bibitem{HS}
G.T. Horowitz and A. Strominger,
{\it Nucl. Phys.} {\bf B 360}, 197 (1991).

\bibitem{Guv}
R. G\"{u}ven, {\it Phys. Lett.} {\bf B 276}, 49 (1992);
{\it Phys. Lett.} {\bf B 212}, 277 (1988).

\bibitem{KLO}
R. Kallosh, A. Linde, T. Ortin, A. Peet and A. van Proeyen,
{\it Phys. Rev.} {\bf D 46},  5278 (1992).

\bibitem{LPSS}
H. L\"u, C.N. Pope, E. Sezgin and K. Stelle,
{\it Nucl. Phys.} {\bf B 456}, 669 (1995).

\bibitem{Str}
A. Strominger, {\it Phys. Lett. } {\bf B 383}, 44 (1996);
hep-th/9512059.
\bibitem{To}
P.K. Townsend, {\it Phys. Lett. } {\bf B 373}, 68 (1996);
hep-th/9512062.
\bibitem{Ts1A}
A.A. Tseytlin, {\it Nucl. Phys.} {\bf B 487}, 141 (1997);
hep-th/9609212.

\bibitem{Ts2}
A.A. Tseytlin,
{\it Mod. Phys. Lett.} {\bf A11}, 689 (1996); hep-th/9601177.
\bibitem{PT}
G. Papadopoulos and P.K. Townsend,
{\it  Phys. Lett.} {\bf B 380}, 273 (1996).
\bibitem{Ts1}
A.A. Tseytlin, Harmonic Superpositions of M-branes,
hep-th/9604035; {\it Nucl. Phys.} {\bf B 475}, 149 (1996).
\bibitem{GKT}
J.P. Gauntlett,  D.A. Kastor, and J. Traschen,
Overlapping Branes in M-Theory, hep-th/9604179; {\it Nucl. Phys.}
{\bf B 478}, 544 (1996).
\bibitem{KKLP}
N. Khviengia, Z. Khviengia, H. L\"u, C.N. Pope,
Intersecting M-Branes and Bound States, hep-th/9605082.

\bibitem{LPX}
H. L\"u, C.N. Pope, and K.W. Xu, Liouville and Toda Solitons in
M-Theory, hep-th/9604058.

\bibitem{CT}
M. Cvetic and A. Tseytlin, {\it Nucl. Phys.} {\bf B 478}, 181 (1996).
\bibitem{KT}
I.R. Klebanov and A.A. Tseytlin, Intersecting $M$-branes as
Four-Dimensional Black Holes, {\it Preprint} PUPT-1616,
Imperial/TP/95-96/41,  hep-th/9604166; {\it Nucl. Phys.} {\bf B 475},
164 (1996).
\bibitem{OS}
N. Ohta and T. Shimizu, Non-extreme Black Holes from Intersecting
M-branes, hep-th/9701095.

\bibitem{LPS2}
H. L\"u, C.N. Pope, and K.S. Stelle,
Vertical Versus Diagonal
Reduction for p-Branes,  hep-th/9605082.
\bibitem{BKO}
E. Bergshoeff, R. Kallosh and T. Ortin,
Stationary Axion/Dilaton Solutions and Supersymmetry,
hep-th/9605059;
{\it Nucl. Phys.} {\bf B 478}, 156 (1996).
\bibitem{CGa}
G. Cl\'ement and D.V. Gal'tsov,
Stationary BPS solutions to dilaton-axion gravity
{\it Preprint} GCR-96/07/02 DTP-MSU/96-11,
hep-th/9607043.

\bibitem{V}
A. Volovich, Three-block p-branes in various dimensions,
hep-th/9608095.
\bibitem{AV}
I.Ya. Aref'eva and A.I. Volovich,
Composite $p$-branes in Diverse Dimensions,
{\it Preprint} SMI-19-96,
hep-th/9611026; {\it Class. Quantum Grav.} 14 (11), 2990 (1997).

\bibitem{AVV}
I.Ya. Aref'eva, K. Viswanathan, A.I. Volovich and I.V. Volovich,
p-Brane Solutions in Diverse Dimensions, hep-th/9701092.
\bibitem{KKP}
N. Khviengia, Z. Khviengia, H. L\"u and  C.N. Pope,
Toward Field Theory of F-Theory, hep-th/9703012.

\bibitem{IM0}
V.D. Ivashchuk and V.N. Melnikov,
Intersecting p-Brane Solutions in Multidimensional Gravity and M-Theory,
hep-th/9612089; {\it Grav. and Cosmol.} {\bf 2}, No 4, 297 (1996).
\bibitem{IM}
V.D. Ivashchuk and V.N. Melnikov,
{\it Phys. Lett. B}  {\bf 403}, 23 (1997).
\bibitem{IMC}
V.D. Ivashchuk and V.N. Melnikov,
Sigma-Model for Generalized  Composite p-branes, hep-th/9705036;
{\it Class. and Quant. Grav.} {\bf 14}, 11, 3001 (1997).

\bibitem{IMR}
V.D. Ivashchuk, M. Rainer and V.N. Melnikov,
Multidimensional Sigma-Models with Composite Electric p-branes,
gr-qc/9705005; {\it Gravit. and Cosm.} {\bf 4}, No1 (13) (1998).

\bibitem{BREJS}
E. Bergshoeff, M. de Roo, E. Eyras, B. Janssen and J.P. van der Schaar,
hep-th/9612095.

\bibitem{AR}
I.Ya. Aref'eva and O.A. Rytchkov,
Incidence Matrix Description of Intersecting p-brane Solutions,
hep-th/9612236.

\bibitem{AEH}
R. Argurio, F. Englert and L. Hourant, Intersection Rules for
$p$-branes, hep-th/9701042.

\bibitem{AIR}
I.Ya. Aref'eva M.G. Ivanov and O.A. Rytchkov,
Properties of Intersecting p-branes in Various Dimensions, hep-th/9702077.

\bibitem{AIV}
I.Ya. Aref'eva, M.G. Ivanov and I.V. Volovich,
Non-Extremal Intersecting p-Branes in Various Dimensions, hep-th/9702079;
{\it Phys. Lett. } {\bf B 406}, 44 (1997).
\bibitem{Oh}
N. Ohta, Intersection Rules for Non-extreme p-branes, hep-th/9702164.
\bibitem{BIM}
K.A. Bronnikov, V.D. Ivashchuk and V.N. Melnikov,
The Reissner-Nordstr\"om Problem for Intersecting Electric and Magnetic
p-Branes, gr-qc/9710054;
{\it Grav. and Cosmol.\/} {\bf 3}, No 3 (11), 203 (1997).

\bibitem{BKR}
K.A. Bronnikov, U. Kasper and M. Rainer,
Intersecting Electric and Magnetic $p$-Branes: Spherically Symmetric
Solutions, gr-qc/9708058.
\bibitem{BGIM}
K.A. Bronnikov, M.A. Grebeniuk, V.D. Ivashchuk and V.N. Melnikov,
Integrable Multidimensional Cosmology for
Intersecting $p$-branes,
{\it Grav. and Cosmol. } {\bf  3}, No 2(10), 105 (1997).
\bibitem{GrIM}
M.A. Grebeniuk, V.D. Ivashchuk and V.N. Melnikov,
Integrable Multidimensional Quantum Cosmology  for Intersecting p-Branes,
{\it Grav. and Cosmol.\/} {\bf 3}, No 3 (11), 243 (1997), gr-qc/9708031.
\bibitem{LMMP}
H. L\"u, J. Maharana, S. Mukherji  and C.N. Pope,
Cosmological Solutions, p-branes and the Wheeler De Witt
Equation,   hep-th/9707182.
\bibitem{LMPX}
H. L\"u, S. Mukherji, C.N. Pope and K.-W. Xu,
Cosmological Solutions in String Theories,  hep-th/9610107.
\bibitem{W}
S. Weinberg, {\it Rev. Mod. Phys.} {\bf 61}, 1 (1989).

\bibitem{IMAJ}
V.D. Ivashchuk and V.N. Melnikov, Multidimensional Ouantum Cosmology with
Intersecting p-branes, {\it Hadronic J.} {\bf 21}, 319 (1998).

\bibitem{Cosm}
M.A. Grebeniuk, V.D. Ivashchuk and V.N. Melnikov, Multidimensional Cosmology
for Intersecting p-branes with Static Internal Spaces, {\it Grav. and
Cosm.}, {\bf 4}, No 2(14) (1998).

\bibitem{MP}
S.D. Majumdar, {\it Phys. Rev. } {\bf 72}, 930 (1947);  \\
A. Papapetrou,  {\it Proc. R. Irish Acad. } {\bf A51}, 191 (1947).

\bibitem{Br}
N.M. Bocharova, K.A. Bronnikov and V.N. Melnikov,
{\it Vestnik MGU (Moscow Univ.)}, {\bf 6}, 706 (1970)(in Russian)
- first MP-type solution with conformal scalar field; \\
K.A. Bronnikov,
{\it Acta Phys. Polonica }, {\bf B4}, 251 (1973);\\
K.A. Bronnikov and V.N. Melnikov,  in
{\it Problems of Theory of Gravitation and Elementary Particles
}, {\bf 5}, 80 (1974) (in Russian)
- first MP-type solution with conformal scalar
and electromagnetic fields.

\bibitem{Sz2}
M. Szyd\l owski, {\it Acta Cosmologica} {\bf 18}, 85 (1992).

\bibitem{GH}
G.W. Gibbons and S.W. Hawking, {\it Phys. Rev.} {\bf D 15}, 2752 (1977).

\bibitem{IM6I}
V.D. Ivashchuk and V.N. Melnikov,
{\it  Int. J. Mod. Phys. D} {\bf 4}, 167 (1995).

\bibitem{IMB}
V.D. Ivashchuk and V.N. Melnikov,
On Singular Solutions in Multidimensional
Gravity, hep-th/9612089;
{\it Grav. and Cosmol.} {\bf 1}, No 3, 204 (1996).

\bibitem{IMJ}
V.D. Ivashchuk and V.N. Melnikov,
Multidimensional Classical and Quantum Cosmology
with Intersecting $p$-branes, hep-th/9708157;
{\it J. Math. Phys.}, {\bf 39}, 2866 (1998).

\bibitem{Br1}
K.A. Bronnikov, Block-orthogonal Brane systems, Black
Holes and Wormholes, hep-th/9710207;
{\it Grav. and Cosmol.} {\bf 4}, No 2 (14), (1998).

\bibitem{IMBl}
V.D. Ivashchuk and V.N. Melnikov,
Mudjumdar-Papapetrou Type Solutions in Sigma-model
and Intersecting $p$-branes, hep-th/9802121,
{\it Class. Quantum Grav.} {\bf 16}, 849 (1999).

\bibitem{I2}
V.D.Ivashchuk, Composite p-branes on Product of Einstein spaces,
{\it Phys. Lett.} {\bf B 434}, 28 (1998).

\bibitem{Dam}
T. Damour, "Gravitation, Experiment and Cosmology", gr-qc/9606079.

\bibitem{Re}
R.D. Reasenberg et. al., {\it Astrophys. J.} {\bf 234}, L219 (1979).
\bibitem{Di}
J.O. Dickey et al., {\it Science} {\bf 265}, 482 (1994).
\bibitem{N}
K. Nordtvedt, {\it Phys. Rev. } {\bf 169}, 1017 (1968).



\end{thebibliography}
\end{document}